\pdfoutput=1
\documentclass[11pt]{revtex4} \usepackage[pdftex]{graphicx}
\usepackage{amsmath} \usepackage{amssymb} \usepackage{amsxtra}

\begin{document}

\title{The newest experimental data for the quarks mixing matrix are in better
agreement with the {\it spin-charge-family} theory predictions than the old ones.
 }
\author{G. Bregar, N.S. Manko\v c Bor\v stnik \\
 Department of Physics, FMF, University of Ljubljana, \\
Jadranska 19, SI-1000 Ljubljana, Slovenia}
\begin{abstract} The {\it spin-charge-family} theory~\cite{NBled2013,NBled2012,norma92,%
norma93,norma94,pikanorma,JMP,norma95,gmdn07,gn,gn2013,NPLB,N2014scalarprop,%
norma2014MatterAntimatter} predicts before the electroweak break four - rather than the 
observed three - massless families of quarks and leptons. The $4 \times 4$ mass matrices of
all the family members demonstrate in this theory the same symmetry, which is determined by
the scalar fields: the two $SU(2)$ triplets (the gauge fields of the family groups) and the three 
singlets, the gauge fields of the three charges ($Q, Q'$ and $Y'$) distinguishing among family 
members. All the scalars have, with respect to the weak and the hyper charge, the quantum 
numbers of the {\it standard model} scalar Higgs~\cite{N2014scalarprop}: $\pm \frac{1}{2}$
 and $ \mp \frac{1}{2}$, respectively. 
Respecting by the {\it spin-charge-family} theory proposed symmetry of mass matrices and 
assuming (due to not yet accurate enough experimental data) that the mass matrices are
hermitian and real, 
we fit the six free parameters of each family member mass matrix to the
experimental data of twice three measured masses of quarks and to the measured quark  
mixing matrix elements, within the experimental accuracy. Since any $3 \times 3$ submatrix
of the $4\times 4$ unitary matrix determines the whole $4\times 4$ matrix uniquely, we are 
able to predict the properties of the fourth family members provided that the experimental 
data are enough accurate, which is not yet the case. We, however, found out that the new
experimental data~\cite{datanew} for quarks fit better to the required symmetry of mass
matrices than the old data~\cite{dataold} and we predict towards which value
will more accurately measured matrix elements move.
The present accuracy of the experimental data for leptons does not enable us to make sensible 
predictions.
\end{abstract}
\maketitle

 \section{Introduction} 
 \label{introduction}

There are several attempts in the literature to reconstruct mass matrices of quarks and leptons 
out of the observed masses and mixing matrices in order to learn more about properties of the 
fermion families~\cite{FRI,FRI1,FRI2,FRI3,FRI4,FRI5,FRI6,FRI7,FRI8,FRI9,FRI10,FRI11}. The
most popular is the $n\times n$ matrix, 
close to the democratic one, predicting that $(n-1)$ families must be very light in comparison with
the $n^{\rm th}$ one. 
Most of attempts treat neutrinos differently than the other family members, introducing the 
Majorana part and the "sea-saw" mechanism. Most often are the number of families taken to
be equal to the number of the so far observed families, while symmetries of mass matrices are 
chosen in several different ways~\cite{AstriAndOthers,AstriAndOthers1,Astri,lugri}. Also 
possibilities with four families are discussed~\cite{four, four1,four2}.

In this paper we follow the requirements of the {\it spin-charge-family}
theory~\cite{NBled2013,NBled2012,norma92,%
norma93,norma94,pikanorma,JMP,norma95,gmdn07,gn,gn2013,N2014scalarprop,%
norma2014MatterAntimatter}, which predicts four coupled families of quarks and leptons and the
mass matrix symmetry, which is the same for all the family members.

The mass matrix of each family member is in the {\it spin-charge-family} theory determined by 
the scalar fields, which carry besides by the {\it standard model} required weak and hyper
charges~\cite{N2014scalarprop} ($\pm \frac{1}{2}$ and $ \mp \frac{1}{2}$, respectively) also
the additional charges: There are two $SU(2)$ triplets, the gauge fields of the family groups, and
three singlets, the gauge fields of the three charges ($Q, Q'$ and $Y'$), which distinguish among 
family members. These scalar fields cause, after getting nonzero vacuum expectation 
values~\cite{N2014scalarprop}, the electroweak break. Assuming that the contributions of all the 
scalar (and in loop corrections also of other) fields to mass matrices of fermions are real and
symmetric, we are left with the following symmetry of mass matrices 
\begin{equation}
 \label{M0}
{\cal M}^{\alpha} = 
\begin{pmatrix} 
- a_1 - a & e & d & b\\ 
e & - a_2 - a & b & d\\ 
d & b & a_2 - a & e \\ 
b & d & e & a_1 - a
 \end{pmatrix}^{\alpha}\,, 
 \end{equation}
the same for all the family members $\alpha\in \{u,d,\nu,e \}$. In App.~\ref{M0SCFT} the 
evaluation of this mass matrix is presented and the symmetry commented. The symmetry of the
mass matrix~Eq.(\ref{M0}) is kept in all loop corrections.
A change of phases of the left handed and the right handed basis - there are ($2n-1$) free choices - 
manifests in a change of phases of mass matrices. We made a choice of the simplest phases.

The differences in the properties of the family members originate in different charges of the family 
members and correspondingly in the different couplings to the corresponding scalar and gauge 
fields~\cite{JMP}.

We fit (sect.~\ref{numericalresultsexp}) the mass matrix elements of Eq.~(\ref{M0}) with $6$ 
free parameters for any family member to the so far measured properties of quarks and leptons 
within the experimental accuracy. That is: {\it For a pair of either quarks or leptons, we fit twice
$6$ free parameters of the two mass matrices to twice three so far measured masses and to 
the corresponding mixing matrix}.

Since we have the same number of free parameters ($6$ parameters determine in the 
{\it spin-charge-family} theory the mass matrix of any family member after the mass matrices 
are assumed to be real) as there are measured quantities for either quarks or leptons (two times 
$3$ masses and $6$ angles of the orthogonal mixing matrix under the simplification that the 
mixing matrix is real and hermitian), we should predict the fourth family masses and the missing 
mixing matrix elements ($V_{u_{i} d_{4}}$, $V_{u_{4} d_{i}}\,, i \in(1,2,3)$) uniquely, provided
that the measured quantities are accurate. The $(n-1)$ submatrix of any unitary matrix determines 
the unitary matrix uniquely for $n\ge 4$. The experimental inaccuracy, in particular for leptons but 
also for the matrix elements of the mixing matrix of quarks, is too large to allow us to estimate the
fourth family masses even for quarks better than very roughly, if at all.

 {\it Yet it turns out that our fitting twice $6$ free parameters to the last  quarks experimental
 data}~\cite{datanew} {\it leads to better agreement with the data than when we fit to the older 
 data}~\cite{dataold}.  
 
We predict correspondingly - taking into account new experimental data and the symmetry of the 
mass matrices of Eq.~(\ref{M0}) - how will the quarks $3 \times 3$ (sub)matrix (of the 
$4 \times 4$) mixing matrix change in next more accurate measurements. 

In our fitting procedure we take into account also the estimations of the influence of the fourth family 
masses to the decays of mesons from Refs.~\cite{vysotsky}, making also our own estimations (pretty
roughly so far, this work is not presented in this paper)~\footnote{M.I. Vysotsky and A. Lenz 
comment in their papers~\cite{vysotsky} that  the fourth family is excluded provided that one 
assumes the validity of the {\it standard model} with one scalar field (the scalar Higgs) while 
extending the number of families from three to four when, in loop corrections, evaluating the 
decay properties of the scalar Higgs. We have, however, several scalars: Two times three triplets 
with respect to the family quantum numbers and three singlets, which distinguish among the family
members~\cite{N2014scalarprop}, all these scalars carry the weak and the hyper charge as the 
scalar Higgs. These scalar fields determine all the masses and the mixing matrices of quarks and
leptons and of the weak gauge fields, what in the {\it standard model} is achieved by the choice of
the scalar Higgs properties and the Yukawa couplings. Our rough estimations of the decay properties
of mesons show that the fourth family quarks might have masses close to $1$ TeV or above.}.

The fact that the new experimental data for quarks fit the symmetry of the mass matrices better
than the old ones might be, together with other predictions of this 
theory~\cite{norma2014MatterAntimatter, N2014scalarprop,NBled2013,JMP}, a promising signal 
that the {\it spin-charge-family} theory is the right step beyond the {\it standard model} (although 
we had to assume in this calculations, due to not yet accurate enough quarks mixing matrix,  that
the mass matrices (Eq.~(\ref{M0}))  are real).

In the {\it spin-charge-family} theory all the family members, the quarks and the leptons,  are treated 
equivalently.  However, the experimental data for leptons are so far too inaccurate to allow us to make 
any accurate enough predictions.

In Sect.~\ref{procedure} the variational procedure to fit the free parameters of the mass matrices 
(Eq.~(\ref{M0})) to the experimental data is discussed. In sect.~\ref{numericalresults} the numerical 
results obtained in the fitting procedure of the free parameters of the mass matrices to two kinds of
the experimental data~\cite{dataold,datanew} are presented. We present the mass matrices for 
quarks, their diagonal values and the mixing matrix, and we present how good our variational procedure 
works. The results are commented and predictions made. Discussions are made in 
Sect.~\ref{discussions}.

In App.~\ref{scft} we offer a very brief introduction into the {\it spin-charge-family} theory, which 
the reader, accepting the proposed symmetry of mass matrices without being curious about the origin 
of this symmetry, can skip. The rest of appendices are meant as the pedagogical addition.

 \section{Procedure used to fit free parameters of mass matrices to
experimental data}
 \label{procedure} 
 %


Mass matrices~Eq.(\ref{M0}), following from the {\it spin-charge-family} theory, are not in 
general real (App.~\ref{nonhermitean}). We, however, assume in this study, due to the
experimental inaccuracy of even the quarks mixing matrix, which does not allow us to extract 
three complex phases, that the mass matrix of any family member (of $u$ and $d$ quarks 
and $\nu$ and $e$ leptons) is real and correspondingly symmetric. We choose the simplest 
phases for the basic states, as discussed in App.~\ref{M0SCFT}~\footnote{In Ref.~\cite{gmdn07}
we made a similar assumption, 
except that we allow there that the symmetry of mass matrices, manifesting  on the tree
level, might be changed in loop corrections. We got in that study dependence of mass matrices
and correspondingly mixing matrices on masses of the fourth family members. The study of 
the symmetry of mass matrices in loop corrections in the massless basis shows that loop 
corrections keep the symmetry, determined by the group content $SU(2) \times SU(2) \times
U(1)$  in all orders~\cite{AN}.}.

The matrix elements of mass matrices, with the loop corrections in all orders taken into
account and manifesting the symmetry of Eq.~(\ref{M0}), are in this paper taken as free 
parameters. Due to this symmetry, required by the family quantum numbers of the scalar
fields~\cite{N2014scalarprop}, there are $6$ parameters having $(n-1)\cdot (n-2)/2$ 
complex phases. Assuming - simplifying the calculations in accordance with the experimental 
inaccuracy,  which does not allow us to take into account that mass matrices might be complex 
and the corresponding mixing matrices unitary - that mass matrices are real and mixing matrices
are correspondingly orthogonal, there are $6$ free real parameters for the mass matrix of any
family member - for $u$ and $d$ quarks and for $\nu$ and $e$ leptons. 

Let us first briefly overview properties of mixing matrices, a more detailed explanation of which 
can be found in App.~\ref{M0SCFT}.

Let $M^{\alpha}$, $\alpha$ denotes the family member ($\alpha=u,d,\nu,e$), be the mass 
matrix in the massless basis (with all loop corrections taken into account). Let $V_{\alpha \beta}$
$= S^{\alpha} S^{\beta \dagger}$, where $\alpha$ represents either the $u$-quark and $\beta$
the $d$-quark, or $\alpha$ represents the $\nu$-lepton and $\beta$ the $e$-lepton, denotes a 
(in general unitary) mixing matrix of a particular pair: the quarks one or the leptons one.
  
 For $n\times n$ matrix ($n=4$ in our case) it follows: \\ 
i. Known matrix elements of the submatrix $(n-1) \times (n-1)$ of an unitary matrix $n \times n$, 
$n\ge 4$ determine the whole unitary matrix $n \times n$ uniquely: The $n^2$ unitarity
conditions determine ($2(2(n-1) +1)$) real unknowns completely. If the sub
matrix $(n-1) \times (n-1)$ of an unitary matrix is made unitary by itself, then
we loose the information of the last row and last column.\\ 
ii. If the mixing matrix is assumed to be orthogonal, then the $(n-1) \times (n-1)$ submatrix 
contains all the information about the $n\times n $ orthogonal matrix to which it belongs and the 
$n(n+1)/2 $ conditions determine the $2(n-1)+1$ real unknowns completely for any $n$. If the 
submatrix of the orthogonal matrix is made orthogonal by itself, then we loose all the information
of the last row and last column. 

In what follows we present the procedure used in our study and repeat the assumptions. 
\begin{enumerate}
\item If the mass matrix $ M^{\alpha}$ is hermitian, then the unitary matrices $S^{\alpha}$ and 
$T^{\alpha}$, introduced in appendix~\ref{nonhermitean} to diagonalize a non hermitian mass 
matrix, differ only in phase factors depending on phases of basic vectors and manifesting in two 
diagonal matrices, $F^{\alpha \,S}$ and $ F^{\alpha\,T}$, corresponding to the left handed and 
the right handed basis, respectively. For hermitian mass matrices we therefore have: $T^{\alpha}$ 
$=S^{\alpha}\, F^{\alpha \,S} F^{\alpha\,T\, \dagger}$. By changing phases of basic vectors
we can change phases of $(2n-1)$ matrix elements.  
\item We take for  quarks and leptons twice three out of twice four diagonal values of the diagonal  
matrix ${\cal M}^{\alpha}_{d}$ and the corresponding mixing matrix $V_{\alpha \beta}$ from
the available experimental data. The fourth family members properties are to be determined, 
together with the corresponding mixing matrix elements,  from the numerical procedure. Each 
mass matrix $ M^{\alpha}$, Eq.~(\ref{M0}), has, if it is real, $6$ free real parameters $
(a^{\alpha}, a^{\alpha}_1,a^{\alpha}_2, b^{\alpha}, e^{\alpha}, d^{\alpha})$, $\alpha$ 
$= (u,d,\nu,e)$.
\item We limit the number of free parameters of the mass matrix of each family member 
$\alpha$ by taking into account $n$ relations among free parameters, in our case $n=4$,
determined by the invariants
\begin{eqnarray} 
\label{invariants}
I^{\alpha}_{1} &=& - \sum_{i=1,4} \, m^{\alpha}_i, \;\quad I^{\alpha}_{2} =
\sum_{i>j=1,4} \, m^{\alpha}_i \, m^{\alpha}_j,\;\nonumber\\ 
I^{\alpha}_{3} &=& -\sum_{i>j>k=1,4} \, m^{\alpha}_i \, m^{\alpha}_j\, m^{\alpha}_k,\;\quad
I^{\alpha}_{4} = m^{\alpha}_1\, m^{\alpha}_2 \, m^{\alpha}_3\, m^{\alpha}_4\,,\nonumber\\
\alpha &=&u,d,\nu,e\,, 
\end{eqnarray} 
which are the expressions appearing at powers of $\lambda_{\alpha}$, $\lambda_{\alpha}^4+ $
$\lambda_{\alpha}^3 I_{1} +$ $\lambda_{\alpha}^{2} I_{2} + $ $\lambda_{\alpha}^1 I_{3} + $
$\lambda_{\alpha}^0 I_{4}=0$, in the eigenvalue equation. The invariants are fixed (within the 
experimental inaccuracy of the data) by the observed masses of quarks and leptons and by the 
chosen value of the fourth family mass $m^{\alpha}_4$. One can express the four invariants with
the parameters of the mass matrix (Eq.~(\ref{M0})): 
 \begin{eqnarray}
\label{osnovneenacbe}
a^{\alpha} &=& \frac{I^{\alpha}_{1}}{4}\,, \nonumber\\
-I^{\alpha}_{2} + 6\, (a^{\alpha})^2 &=& (a^{\alpha}_1)^2 + (a^{\alpha}_2)^2 + 2\, (b^{\alpha})^2
+ 2\, (d^{\alpha})^2 + 2\, (e^{\alpha})^2\,, \nonumber\\
I^{\alpha}_{3} - 2 \,a^{\alpha}\, I^{\alpha}_2 + 8 \,(a^{\alpha})^3 &=& - 8 \,
d^{\alpha} \, e^{\alpha} \, b^{\alpha}\,, \nonumber\\ 
I^{\alpha}_{4} - a^{\alpha}\, I^{\alpha}_3 + (a^{\alpha})^2\, I^{\alpha}_2 - 3\, (a^{\alpha})^4 &=&
((a^{\alpha}_1)^2 + (b^{\alpha})^2)\, ((b^{\alpha})^2 + (a^{\alpha}_2)^2) - 2\, a^{\alpha}_1\,
a^{\alpha}_2 \, ((e^{\alpha})^2 - (d^{\alpha})^2) - \nonumber\\
&& 2\, (b^{\alpha})^2\, ((e^{\alpha})^2 + (d^{\alpha})^2) + ((e^{\alpha})^2 -
(d^{\alpha})^2)^2\,.
\end{eqnarray} 
Correspondingly there are ($6-4$) free real parameters left for each mass matrix, after a choice is
made for the mass of the fourth family member. 
\item
The diagonalizing matrices $S^{\alpha}$ and $S^{\beta}$, each depending on ($6-4$) free real 
parameters, after the choice of the fourth family masses, 
are for real and symmetric mass matrices orthogonal. They follow from the procedure 
\begin{eqnarray} 
\label{mdiag} 
M^{\alpha}&=& S^{\alpha} \,{\bf M}^{\alpha}_{d}\,T^{\alpha \, \dagger}\,, \quad
T^{\alpha}=S^{\alpha}\, F^{\alpha \,S} F^{\alpha\,T\, \dagger}\,,\nonumber\\
{\bf M}^{\alpha}_{d}&=& (m^{\alpha}_1, m^{\alpha}_2, m^{\alpha}_3,
m^{\alpha}_4)\,, 
\end{eqnarray}
provided that $S^{\alpha}$ and $S^{\beta}$ fit the experimentally observed mixing matrices 
$V^{\dagger}_{\alpha \beta}$ within the experimental accuracy and that $ M^{\alpha}$ and 
$ M^{\beta}$ manifest the symmetry presented in Eq.~(\ref{M0}). We keep the symmetry
of the mass matrices accurate. One can proceed in two ways.
\begin{eqnarray}
\label{SVud}
A.:\;\;S^{\beta} = V^{\dagger}_{\alpha \beta} S^{\alpha}\,\,,&&\quad
B.:\;\;S^{\alpha } = V_{\alpha \beta} S^{\beta}\,, \nonumber\\
A.:\;\;V^{\dagger}_{\alpha \beta}\,S^{\alpha }\, {\bf M}^{\beta }_{d}
\,S^{\alpha \dagger } V_{\alpha \beta}= M^{\beta} \,\,,&&\quad B.:\;\;V_{\alpha
\beta}\,S^{\beta }\, {\bf M}_{d}^{\alpha} \,S^{\beta \dagger }
V^{\dagger}_{\alpha \beta}= M^{\alpha} \,.
\end{eqnarray}
The indices $\alpha$ and $\beta$ determine to which family member the matrix $S$ 
corresponds, or to which two family members the mixing matrix $V$ corresponds. In the
case $A.\,$ one obtains from Eq.~(\ref{mdiag}), after requiring that the mass
matrix $M^{\alpha}$ has the desired symmetry,  
the matrix $S^{\alpha}$ and the mass matrix $M^{\alpha}$ ($=S^{\alpha }\,
{\bf M}^{\alpha}_{d} \,S^{\alpha \dagger }$), from where we get the mass matrix 
$M^{\beta}$ $= V^{\dagger}_{\alpha \beta}\,S^{\alpha }\, {\bf M}^{\beta }_{d} \,
S^{\alpha \dagger } V_{\alpha \beta}$. 
In case $B.\,$ one obtains equivalently 
the matrix $S^{\beta}$, from where we get $M^{\alpha}$ ($=V_{\alpha \beta}\,S^{\beta }\,
{\bf M}_{d}^{\alpha} \,S^{\beta \dagger } V^{\dagger}_{\alpha \beta}$). We use both ways
iteratively taking into account the experimental accuracy of masses and mixing matrices. 
\item Under the assumption of the present study that the mass matrices 
are real, form the corresponding orthogonal diagonalizing matrices $S^{\alpha}$ and $S^{\beta}$
the orthogonal mixing matrix $V_{\alpha \beta}$, which depends on at most $6\,(=\frac{n(n-1)}{2})$
free real parameters (App.~\ref{nonhermitean}).\\
We should not forget, that the assumption of the real and symmetric mass matrices, leading to 
orthogonal mixing matrices, might not be an acceptable simplification. However, the experimental
inaccuracy in particular of the mixing matrices does not allow us (yet) to take complex phases 
into account. (In the next step of study, with hopefully more accurate experimental data, we 
will relax conditions on hermiticity of mass matrices and correspondingly on orthogonality of mixing 
matrices, allowing them to be unitary.)  We expect and shall find that too large experimental
inaccuracy leaves the fourth family masses in the present study quite undetermined, even for 
quarks.
\item We study quarks and leptons equivalently. 
\item The difference among family members originate in the eigenvalues of the operators 
$(Q^{\alpha},Q'^{\alpha},Y'^{\alpha})$, which in loop corrections in all orders contribute to all 
mass matrix elements, causing the difference among family members~\footnote{There are also
Majorana like terms contributing in higher order loop corrections~\cite{JMP}, which might strongly
influence in particular the neutrino mass matrix.}.
\end{enumerate} 

Let us summarize. If the mass matrix of a family member obeys the symmetry required by the 
{\it spin-charge-family} theory, which in a simplified version is assumed to be real and symmetric 
 (as it is taken in this study), the matrix elements of the mixing matrices of either quarks or leptons
 are correspondingly real, each of them with $\frac{n(n-1)}{2} =6$ parameters. The  $2 \times 6$ 
free parameters of the two mass matrices (of either the quark or the lepton pair) are in this case 
determined by $2 \times 3$ measured masses and the $6$ parameters of the  mixing matrix of the
corresponding pair. The accurate enough experimental data  would therefore determine the fourth 
family masses of each family member and the mixing matrix elements of the fourth families 
members for each pair.

Since the so far measured masses and in particular the measured mixing matrices are not 
determined accurately enough, 
we can in the best case expect that the masses and the mixing matrix elements of the fourth 
families will be determined only within some (quite large) intervals.
 
In the Subsect.~\ref{reductionofpar} we present in more details the procedure used to determine 
free parameters  of mass matrices by taking into account the experimental inaccuracy.

\subsection{Free parameters of mass matrices after taking into account invariants} 
\label{reductionofpar} 

We present in this subsection the numerical procedure used in our calculations for fitting free
parameters of each family member mass matrix ($a^{\alpha}$, $a^{\alpha}_1$, 
$a^{\alpha}_2$, $b^{\alpha}$, $e^{\alpha}$, $d^{\alpha}$; $\alpha = (u,d,\nu,e)$),
 Eq.~(\ref{M0}), to the experimental data: $2\times 3$ masses and the $3\times 3$ 
submatrix of the corresponding $4 \times 4$ unitary mixing matrix for either quarks or leptons.

We make the assumption that the mass matrices (Eq.~(\ref{M0})) are real, since the elements of
the mixing matrices are not known accurately enough to extract $3$ complex phases of the unitary 
$4 \times 4$ mixing matrix, either for quarks or (even much less) for leptons. Correspondingly are the diagonalizing matrices orthogonal and so are the mixing matrices. 

The accurate $3\times 3$ submatrix of the $4\times 4$ unitary matrix would determine the unitary matrix uniquely. The assumption that the mass matrices (Eq.~(\ref{M0})) are real makes the mixing matrix orthogonal, determined by $6$ parameters. These $6$ parameters and twice $3$ masses of the pair of either quarks or leptons, if accurate, would determine twice $6$ 
parameters of mass matrices uniquely. It is the experimental inaccuracy, which makes this fitting procedure very demanding, in particular since the $3\times 3$ submatrix of the $4 \times 4$ mixing matrix is, within the experimental inaccuracy, close to an unitary matrix.

Our variational procedure must find the best fit to the experimental data keeping the symmetry of 
any of the mass matrices of a pair of quarks or leptons presented in Eq.~(\ref{M0}).
 
The result of the variational procedure are correspondingly $6$ free parameters of each of the two
mass matrices of a pair, those with the best fit to the experimental data within their accuracy - for 
either the mixing matrix or for twice the three masses. The fitting procedure offers correspondingly
intervals for the fourth family masses as well as for the mixing matrix elements of the fourth family 
members to the rest three of the family members.

Although we have investigated leptons and quarks, it has turned out that the experimental data for
leptons are not yet accurate enough that we would be able to make some valuable predictions for 
leptons. We therefore concentrate in this paper on quarks.
   
We skip in this subsection, for the sake of simplicity, the family member index $\alpha$. It appears 
useful, in purpose of numerical evaluation, to take into account for each family member its mass
matrix invariants, Eq.~(\ref{invariants}), which are the coefficients of the characteristic polynomials, 
and to make a choice of the fourth family masses or rather $a=\frac{1}{4} \,I_{1}$ 
(Eq.~(\ref{iprime})) instead of the fourth family mass. We also introduce new parameters $f$, $g$, 
$q$ and $r$ 
\begin{eqnarray}
\label{deqrplusminus} 
a, b\,, \quad f=d+e \,, \quad g=d-e\,,\quad q=\frac{a_1 + a_2}{\sqrt{2}}\,, \quad r=
\frac{a_1 - a_2}{\sqrt{2}}\,,
\end{eqnarray} 
and express the parameters $a$, $q$, $r$, $b$, $f$ and $g$ of the mass matrix, Eq.~(\ref{M0}), 
with invariants, Eq.~(\ref{invariants}), 
\begin{eqnarray} 
\label{iprime} 
a &=& \frac{I_{1}}{4}\,,\nonumber\\ 
I'_{2}&=& -I_{2} + 6a^2 = q^2 + r^2 + 2b^2 + f^{2} + g^{2}\,, \nonumber\\
I'_{3}&=& I_{3}- 2a I_{2}+ 8 a^3 = - 2 b \, (f^{2} - g^{2})\,,\nonumber\\ 
I'_{4}&=& I_{4} - a I_{3} + a^2 I_{2}-3 a^4 \nonumber\\
        &=& \frac{1}{4} (q^2-r^2)^2+ (q^2+r^2) b^2 + \frac{1}{2}\,(q^2-r^2)\cdot 2\, g \,f - b^2\, 
        (f^2 +g^2) + \frac{1}{4}\, (2\, g\, f)^2 +  b^4\,.        
\end{eqnarray}
We do this to reduce the number of free parameters for each mass matrix from $6$ to $3$ by making a choice of (twice) $3$ masses within the experimentally allowed values. It turns out, namely, when testing the stability of the variational procedure against changing masses of the family members within the experimental inaccuracy, that the experimental inaccuracy of twice three known masses of a pair of two family members (either quarks or leptons) influences parameters of the mass matrices and correspondingly the fourth family masses and the corresponding mixing matrix much less than the inaccuracy of the matrix elements of the $3 \times 3$ mixing submatrix. 
We test the stability of the variational procedure against changing masses of the family members within the experimental inaccuracy after each variational procedure.

We eliminate, using the second and the third equation of Eq.~(\ref{iprime}), the parameters $f$ and $g$, expressing them as functions of $I'_{2}$ and $I'_{3}$ and the parameters $a$, $r$, $q$ and $b$ 
\begin{eqnarray} 
\label{fg}
f^2&=&\frac{1}{2} \left(I_2'-q^2-r^2-2b^2-\frac{I_3'}{2 b} \right) \nonumber \\ 
g^2&=& \frac{1}{2} \left(I_2'-q^2-r^2-2b^2+\frac{I_3'}{2 b} \right)\,.
\end{eqnarray} 
To avoid that the product $g f$ appearing in the last equation of Eq.~(\ref{iprime}) would be 
expressed by the square root of the right hand sides of Eq.~(\ref{fg}) we square the last equation 
of Eq.~(\ref{iprime}) (after putting $g\, f$ on one side of the equation and all the rest on the other 
side of the equation). The last four free parameters ($a,b,q,r$) are now related as follows
\begin{eqnarray} 
\label{qr}
&  &\{- \frac{1}{2} (q^4 + r^4) + (-2 b^2 + \frac{1}{2} (I_2'- 2 b^2)) (q^2+r^2) \nonumber\\ 
&+& (I'_{4}-b^4- \frac{1}{4}\, ((I_{2}'-2b^2)^2 - I_3'^2/4b^2) + b^2 (I_{2}' -2b^2))\}^2 
\nonumber\\ 
&=&\frac{1}{4} (q^2-r^2)^2 ( (I_{2}'-2b^2 - (q^2 +r^2))^2 - I_3'^2/4b^2)\,,
\end{eqnarray} 
what reduces the number of free parameters to $3$ for each member of the pair, either of quarks 
or of leptons. Only powers of $q^2$ contribute, and since the term $q^8$ cancels we are left with 
a cubic equation for $q^2$ 
\begin{eqnarray} 
\label{q2} 
\beta\, q^6 + \gamma \,q^4 + \delta\, q^2+ \rho =0\,.
\end{eqnarray}
Coefficients ($\beta,\gamma,\delta, \rho $) depend on the $3$ free remaining parameters
($a,b,r$) and the three (within experimental accuracy) known masses. These twice $3$ free 
parameters for a pair of either quarks or leptons must be determined from the corresponding
measured matrix elements of the $3\times 3$ submatrix of the $4\times4$ mixing matrix
(which is assumed to be orthogonal).

We investigate the concordance between the {\it spin-charge-family}, which defines the 
symmetry of the mass matrix (Eq.~(\ref{M0})) of each family member, and the experimental 
data offering masses of twice three families and, due to the theory, the matrix elements of
the $3\times 3$ submatrices of the two $4\times 4$ mixing matrices.

We proceed as follows: i. First we make a choice of all the masses of $u$ and $d$ quarks -
twice the three known ones within the experimentally allowed values, while the fourth masses 
are taken as free parameters. ii. Then we look for the best fit of twice six mass matrix 
parameters - after taking into account all the relations of Eqs.~(\ref{iprime}, \ref{fg}, 
\ref{qr}, \ref{q2}) - to the six parameters of the mixing matrix, which all are determined only
within experimental accuracy. We repeat the same procedure with several choices of masses: 
for the three known masses we chose values within the experimentally allowed intervals, the 
fourth masses are chosen from $300$ GeV to $1200$ GeV.
The investigations confirm that the experimental uncertainties of the lower three masses of quarks 
influence our fitting procedure very little.

The fixed masses of all quarks simplify calculations substantially. The parameter $a=-(m_1+m_2
+ m_3+m_4)/4$ becomes a constant and the only parameters that remain to be determined 
from the mixing matrix are $b$ and $r$ for the $u$ and the $d$ quarks, that is $4$ parameters. 

For each type of quarks we look for the allowed region in the parameter space of $b$ and $r$ in 
which the remaining three parameters ($f$, $g$ and $q$) solve the equations (\ref{iprime}) as 
real numbers. This condition requires that $f$ and $g$, which follow from Eq.~(\ref{fg}), are
 real and that the solution for $q$ from Eq.~(\ref{qr}) is also real. We end up with the inequality 
\begin{equation}
\label{binterval}
I_{2}' -2\, b^2 - (q^2 + r^2) \geq \left| \frac{I_3'}{2b} \right| \,,
\end{equation} 
which determines the maximal and minimal positive $b$ (both appearing at $q=0=r$). 
Eq.~(\ref{qr}) is insensitive to the sign of $b$. The sign of $f$ and $g$ manifests only in 
Eq.~(\ref{fg}), where the change $b \rightarrow -b$ interchanges $f$ and $g$ 
($f\leftrightarrow g$). Since Eq.~(\ref{qr}) includes only even powers of $q$, $r$ and $b$, the
signs need to be studied in a detailed way.

There are several matrix transformations of the kind $M'=P_i M P_i^T$, $M$ is presented in 
Eq.~(\ref{M0}), $P_{i}$ of which 
can be read from Eq.~(\ref{osnovneenacbe}) 
%
\begin{equation} P_1= \left(
\begin{array}{cccc}
 0 & 0 & 0 & 1 \\ 0 & 0 & 1 & 0 \\ 0 & 1 & 0 & 0 \\ 1 & 0 & 0 & 0 \\ 
\end{array}
\right); \quad
                       P_2= \left(
\begin{array}{cccc}
0 & 1 & 0 & 0\\ 1 & 0 & 0 & 0 \\ 0 & 0 & 0 & 1 \\ 0 & 0 & 1 & 0 \\
\end{array} 
\right); \quad
                       P_3= \left(
\begin{array}{cccc}
-1 & 0 & 0 & 0 \\ 0 & 1 & 0 & 0 \\ 0 & 0 & -1 & 0 \\ 0 & 0 & 0 & 1 \\
\end{array}
\right);\nonumber 
\end{equation}
\begin{equation}
                       P_4= \left( 
\begin{array}{cccc}
-1 & 0 & 0 & 0 \\ 0 & -1 & 0 & 0 \\ 0 & 0 & 1 & 0 \\ 0 & 0 & 0 & 1 \\ 
\end{array}
\right); \quad 
                      P_5= \left( 
\begin{array}{cccc}
-1 & 0 & 0 & 0 \\ 0 & 1 & 0 & 0 \\ 0 & 0 & 1 & 0 \\ 0 & 0 & 0 & -1 \\
\end{array}
\right); \quad 
                       P_6= \left( 
\begin{array}{cccc}
 1 & 0 & 0 & 0 \\ 0 & 0 & 1 & 0 \\ 0 & 1 & 0 & 0 \\ 0 & 0 & 0 & 1 \\
\end{array}
\right);
\nonumber 
\end{equation} 
\begin{equation} 
                      P_7= \left( 
\begin{array}{cccc}
 0 & 0 & 0 & 1 \\ 0 & 1 & 0 & 0 \\ 0 & 0 & 1 & 0 \\ 1 & 0 & 0 & 0 \\
\end{array} \right)\,.
\label{permutations} 
\end{equation} 
These transformations cause changes of the parameters of the mass matrix of
 Eq.~(\ref{M0}), 
and correspondingly also of the new parameters (Eq.~(\ref{deqrplusminus})) as follows 
\begin{eqnarray} 
\label{transformations}
P_1&&: \; a_1 \rightarrow -a_1;\; a_2 \rightarrow -a_2; \quad q \rightarrow
-q;\; r \rightarrow -r; \nonumber \\ 
P_2&&: \; a_{1} \leftrightarrow a_2; \quad
r \rightarrow -r; \nonumber \\ 
P_3&&: \; e \rightarrow -e; \; b \rightarrow -b; \quad f \leftrightarrow g; \nonumber \\ 
P_4&&: \; d \rightarrow -d; \; b \rightarrow -b; \quad f \leftrightarrow -g; \nonumber \\ 
P_5&&: \; e \rightarrow -e; \; d \rightarrow -d; \quad f \rightarrow -f,\; g \rightarrow -g; 
\nonumber\\ 
P_6&&: \; e \leftrightarrow d;\; a_2 \rightarrow -a_2; \quad q \leftrightarrow r;\; 
g \rightarrow -g; \nonumber \\ 
P_7&&: \; e \leftrightarrow d; \, a_1 \rightarrow -a_1; \quad q \leftrightarrow -r; g \rightarrow -g
\, .
\end{eqnarray}

We make use of these transformations in the way that we find the solution of equations 
(\ref{iprime}) for one set of signs of parameters $q$, $r$, $b$, $f$ and $g$ and then obtain all 
the other sets of solutions using transformations (\ref{transformations}).

The procedure regarding the signs is the following: i. We see from Eq.~(\ref{iprime}) that the 
signs of $q$ and $r$ can be chosen arbitrarily. ii. Then we chose $b>0$ and calculate $f^2$ 
and $g^2$ from Eq.~(\ref{fg}). iii. We see in Eq.~(\ref{iprime}) that $f$ and $g$ are present 
only as powers of $f^2$, $g^2$ and $f g$. To obtain the valid sign of $f$ and $g$ we must 
therefore determine only the sign of the product $f g$. iv. The solution for $b<0$  follows
when using  the transformation $P_4$. v. After obtaining in such a procedure all possibilities, 
the region of allowed parameters in the space of parameters $b$ and $r$ follows from 
Eq~(\ref{binterval}).
Since $I_2'$ is positive by definition, we see that the largest value of positive $b$ is obtained when
$q=r=0$. In this case we have the inequality 
\begin{equation} 
\label{b0}
I_{2}' -2 b^2 \geq \left| \frac{I_3'}{2b} \right| \, ,
\end{equation} 
which bounds $b$ from below and from above. The two limits, $b_{min}$ and $b_{max}$, are 
the solutions of the equation which follows from the inequality. We introduce a new parameter 
$\eta_b$
\begin{equation} 
\label{b1} b = b_{min}+ \eta_b (b_{max}-b_{min}) \, . 
\end{equation}
When we fix a certain $b$ with some choice of $\eta_b \in \big[0,1 \big]$ then $r$ and $q$ are 
limited from above, while they are limited from below by $0$. Therefore the largest $r$, 
$r_{max}$, solves the equation 
\begin{equation} 
\label{2}
I_{2}' -2\, b^2 - r_{max}^2= \left| \frac{I_3'}{2b} \right| \, .
\end{equation} 
We introduce a new parameter $\eta_r \in \big[0,1 \big]$
\begin{equation} 
\label{r0} 
r = \eta_r r_{max}\, .
\end{equation} 
For a fixed $r$, $q$ is limited from above by the inequality 
\begin{equation} 
\label{r1} 
I_{2}' -2\, b^2 - (q_{max}^2 + r^2) \geq \left| \frac{I_3'}{2b} \right| \, . 
\end{equation} 
In the space of new parameters $\eta_b$ and $\eta_r$ the cubic equation (\ref{qr}, \ref{q2}) 
selects the region in which a regular solution exists. This region depends on the four masses of each
 family member ($m_i$; $i\in (1,\dots,4)$). 

Introducing the polar coordinates, centered at $\eta_b=1/2, \,\eta_r=1/2$, one finds
that the angle $\gamma$ lies in the interval $\big[\gamma_{min}, \gamma_{max} \big]$, where 
$\gamma_{min}<0$. For each chosen $\gamma$ the radius takes values between $\rho_1 
(\gamma)$ and $\rho_2 (\gamma)$. It is useful to introduce new parameters $\eta_{\gamma}$ 
and $\eta_{\rho}$ as follows 
\begin{eqnarray} 
\label{gammarho} 
\gamma &=& \gamma_{min} + \eta_{\gamma}\, (\gamma_{max} - \gamma_{min})\,, 
\nonumber\\ 
\rho       &=&       \rho_{min} + \eta_{\rho}\, (\rho_{max} - \rho_{min}) \, , 
\end{eqnarray} 
with $\rho_{min}$ and $\rho_{max}$ functionally depending on $\eta_{\gamma}$.

\begin{figure}[h]
\begin{center}
\includegraphics[width=7cm,angle=0]{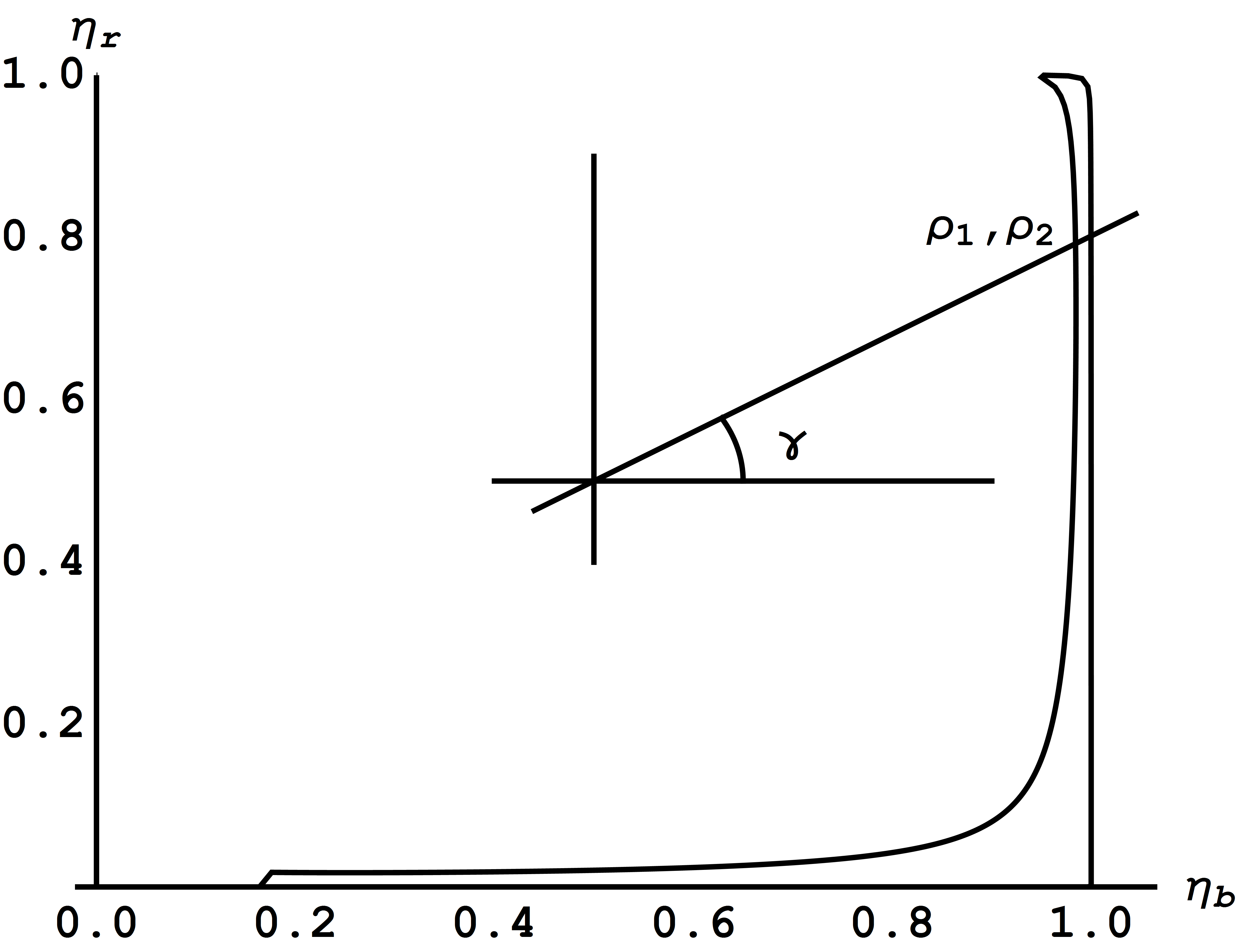}
\end{center}
\caption{The allowed region for the $d$-quarks with masses choice ($m_1^d=2.9, \, m_2^d=
55, \, m_3^d=2900, \, m_4^d=650\,000$) MeV in the space of parameters $\eta_b$, $\eta_r$.}
\label{bumerang}
\end{figure}

An example of such a region for the $d$-quarks with the choice for their masses  $(m_1^d=2.9$ MeV, 
$ m_2^d=55$ MeV,  $m_3^d=2900$ MeV, $ m_4^d=650\,000$ MeV)  is presented on Fig.~\ref{bumerang}.
The allowed region lies between the curve and the two axes.

The calculation reduces, for a pair of quarks or for a pair of leptons, for the chosen values of twice 
four masses of a pair (twice three masses taken within by the experimental data determined 
values, the two fourth ones are taken as parameters) to searching for a minimum of the total discrepancy between the measured mixing matrix elements (with the measuring inaccuracy taken 
into account) and the calculated elements, which depend on $2 \times 2$ parameters, 
($\eta_{\gamma}^u$, $\eta_{\rho}^u$ and $\eta_{\gamma}^d$, $\eta_{\rho}^d$) or 
($\eta_{\gamma}^{\nu}$, $\eta_{\rho}^{\nu}$ and $\eta_{\gamma}^e$, $\eta_{\rho}^e$), 
Eq.~(\ref{sigma}). 
We minimize, for each choice of $2 \times 4$ of masses of a pair, the uncertainty $\sigma_{ud}$ 
(and equivalently for $\sigma_{\nu e}$):
\begin{eqnarray}
\label{sigma} 
\sigma_{ud}\quad &=& \sqrt{\sum_{(i,j)=1}^{3}\, \left( \frac{V_{u_i d_j\, exp} - 
V_{u_i d_j\, cal}}{\sigma_{V_{u_i d_j\, exp}}} \right)^2}\,, \nonumber\\ 
\delta V_{u_i d_j}  &=& \left|\frac{V_{u_i d_j\, exp} - V_{u_i d_j\, cal}}{\sigma_{V_{u_i d_j\, exp}}} \right|\,, 
\end{eqnarray} 
where the two indexes ${}_{exp}$  and ${}_{cal}$ determine experimental and calculated values, 
respectively, and  $\delta_{V_{u_i d_j}}$  determine the discrepancy  between the experimental 
and the calculated values for the 
$3\times 3$ submatrix of the corresponding $4 \times 4$ mixing matrix. 
The experimental values for the quark mixing matrix, the old ones~\cite{dataold} and the new 
ones~\cite{datanew}, are  
presented in Eqs.~(\ref{vudold}, \ref{vudnew}), while the used  twice three quark masses, 
recalculated to the energy scale of the mass of $Z$ boson, $M_{Z}$, are presented in 
Eq.~(\ref{mumd}). 

The search in the 4 dimensional unit cube is performed in the way that we fix the two values of 
parameters $\eta_{\gamma}^u$ and $\eta_{\gamma}^d$ and do a search in the two 
dimensional plane of the other two parameters.

The minimizing process is repeated for several choices of the experimentally allowed values for
twice three measured masses and for several choices of the two fourth family masses in the 
interval $300 $ GeV $\leq (m_{u_4}, m_{d_4})  \leq 1\,200 $ GeV (for leptons we have checked 
the interval $60 $ GeV $\leq (m_{u_4}, m_{d_4})  \leq 1\,000 $ GeV).

The numerical procedure, used in this contribution, is designed for quarks and for lepton, although 
we present in this paper the results for quarks only.

\section{Numerical results} 
\label{numericalresults} 

We present in this paper predictions obtained when taking into account by the 
{\it spin-charge-family} theory required properties of mass matrices (Eq.~(\ref{M0})) and fitting 
free parameters of mass matrices to the measured properties of quarks, using the procedure 
explained in Sects.~\ref{procedure}, \ref{reductionofpar}. Although we have 
performed calculations also for leptons, it turned out that the experimental accuracy of the data 
for leptons are not yet high enough that we could make reliable predictions for them. The 
experimental accuracy is not high enough even for quarks to make use of one complex phase of 
the measured matrix elements of the mixing matrix. We therefore make in this paper the 
assumption that the mass matrices are real. Correspondingly the matrices, which diagonalize
mass matrices, are orthogonal and so is orthogonal also the mixing matrix.

The measured $9$ matrix elements of the quarks mixing matrix form, within the experimental 
inaccuracy, almost unitary $3 \times 3$ matrix. Correspondingly are the results of our fitting 
procedure of twice $6$ parameters of the two $4 \times 4$ mass matrices of Eq.~(\ref{M0}) to 
twice three measured masses and to the measured $9$ matrix elements of the $3 \times 3$ 
submatrix of the corresponding $4 \times 4$ mixing matrix sensitive mostly to the accuracy of 
the measured matrix elements of the mixing matrix.

It turns out, as expected, that inaccuracy of the measured mixing matrix elements does not 
allow us to tell much about the fourth family masses, while inaccuracies of two times three quark 
masses influence the results much less: Quite large intervals for the fourth family masses change 
the calculated  mixing matrix elements of the $3 \times 3$ submatrix very little. The results show
that the most trustworthy might
 be results pushing the fourth family quarks to approximately $1$ TeV or above.

What comes out of our calculations is that the new experimental data~\cite{datanew} for the 
mixing matrix elements, Eq.~(\ref{vudnew}), fit better the predicted symmetry of
Eq.~(\ref{M0}) than the old ones~\cite{dataold}, Eq.~(\ref{vudold}), and that we {\it can 
correspondingly 
predict, following} the {\it spin-charge-family} theory, {\it how will}  {\it the matrix 
elements of the $3 \times 3$ submatrix of the $4 \times 4$ mixing matrix change in next 
measurements}. We predict also the matrix elements of the fourth family members.
We shall see that $V_{u_{i} d_{4}}$ and $V_{u_{4} d_{i}}$, $i\in (1,2,3)$, do not change
very so strongly with the increasing fourth family masses, as one would expect. 
Results of the present paper, together with the results of the previous
works~\cite{norma2014MatterAntimatter,N2014scalarprop}, support the hope that the 
{\it spin-charge-family} theory might be the right next step beyond the {\it standard model}.

Using the procedure, explained in sect.~\ref{procedure} (to some extend presented already in 
Ref.~\cite{gn2013}) we are looking for the $4 \times 4$ in this paper taken to be real and 
correspondingly symmetric mass matrices for quarks, obeying the symmetry of Eq.~(\ref{M0}), 
which reproduce in accordance with Eq.~(\ref{sigma}) as accurately as possible the measured 
properties - masses and mixing matrices - of the so far observed three families of quarks, and 
which are in agreement also with the experimental limits for the appearance of the fourth family 
masses and of the mixing matrix elements to the lower three families, as presented in 
Refs.~\cite{dataold,datanew,vysotsky,four,four1,four2}. (We have made also our own rough 
estimations for the limitations which follow from the meson decays to which the fourth family
 members participate. Our estimations are still in progress.)

A lot of effort was put into the numerical procedure (sect.~\ref{procedure}) to ensure that we fit 
the parameters of mass matrices to the experimental values within the experimental inaccuracy
in the best way, that is with the smallest errors (Eq.~(\ref{sigma})).

The results manifest that the mass matrices are close to the democratic ones, which is, as 
expected, more and more the case the higher might be the fourth family masses, and it is true 
for quarks and leptons.

To test the predicting power of our model in dependence of the experimental inaccuracy of 
masses and mixing matrices, we compare among themselves all the results of the fitting 
procedure, Eq.~(\ref{sigma}), obtained when changing the fourth family quark masses in the
interval of $300$ GeV to $1700$ GeV, for either old~\cite{dataold} or new~\cite{datanew} 
experimental data. 

We look for several properties of the obtained mass matrices:\\
{\bf i.} $\;\;$ We test the influence of the experimentally declared inaccuracy of the
$3 \times 3$ submatrices of the $4 \times 4$ mixing matrices and of the twice $3$ measured masses on the prediction of the fourth family masses.\\
{\bf ii.} $\;\;$ We look for how do the old and the new matrix elements of the measured mixing 
matrix influence the accuracy with which the experimental data are reproduced in the procedure 
which takes into account the symmetry of mass matrices.\\ 
{\bf iii.} $\;\;$ We look for how different choices for the masses of the fourth family members
limit the inaccuracy of particular matrix elements of the mixing matrices or the inaccuracy of the three lower masses of family members.\\ 
{\bf iv.} $\;\;$ We test how close to the democratic mass matrix are the obtained mass matrices
in dependence of the fourth family masses.\\
{\bf v.} $\;\;$ We look for the predictions of the $4 \times 4$ mass matrices with the symmetry 
presented in Eq.~(\ref{M0}).

\subsection{Experimental data used in this calculations  }
\label{numericalresultsexp} 

We take for the quark 
masses  the experimental values~\cite{dataold}, recalculated to the $Z$ boson mass scale.
We take two kinds of the experimental data for the quark mixing matrices with the experimentally
declared inaccuracies for the so far measured $3 \times 3$ mixing matrix elements: The older data 
from~\cite{dataold} and the latest data~\cite{datanew}. We assume, as predicted by the 
{\it spin-charge-family} theory, that these nine matrix elements belong  to the $4 \times 4$
unitary mixing matrix.

We first do the calculations, explained in Sect.~\ref{procedure}, with the older experimental 
data~\cite{dataold}
\begin{equation} 
\label{vudold}
|V_{ud}|= \begin{pmatrix}
 0.97425 \pm 0.00022 & 0.2252 \pm 0.0009 & 0.00415 \pm 0.00049 & |V_{u_1 d_4}|\\
 0.230 \pm 0.011       & 1.006 \pm 0.023     & 0.0409 \pm 0.0011    & |V_{u_2 d_4}|\\ 
 0.0084 \pm 0.0006    & 0.0429 \pm 0.0026  & 0.89 \pm 0.07          & |V_{u_3 d_4}|\\ 
 |V_{u_4 d_1}|          & |V_{u_4 d_2}|         & |V_{u_4 d_3}|         & |V_{u_4 d_4}| \end{pmatrix}\,, 
\end{equation} 
and then we repeat all the calculations also with the new experimental data~\cite{datanew} 
\begin{equation}
\label{vudnew} 
|V_{ud}|= \begin{pmatrix}
 0.97425 \pm 0.00022 & 0.2253 \pm 0.0008   & 0.00413 \pm 0.00049 & |V_{u_1 d_4}|\\ 
 0.225 \pm 0.008       & 0.986 \pm 0.016       & 0.0411  \pm 0.0013   & |V_{u_2 d_4}|\\ 
 0.0084 \pm 0.0006    & 0.0400 \pm 0.0027    & 1.021 \pm 0.032       & |V_{u_3 d_4}|\\ 
|V_{u_4 d_1}|           & |V_{u_4 d_2}|           & |V_{u_4 d_3}|          & |V_{u_4 d_4}|
 \end{pmatrix}\,.
\end{equation} 
For the quark masses at the energy scale of $M_{Z}$ we take 
\begin{eqnarray} 
\label{mumd} 
{\bf M}^{u}_{d}/{\rm MeV/c^2} &=& (1.3 + 0.50 - 0.42 ,\, 619 \pm 84 ,\, 172\,000. \pm 760. ,\,\;
300\,000.\leq m_{u_4}  \leq 1\, 200 \,000.)\,,\nonumber\\ 
{\bf M}^{d}_{d}/{\rm MeV/c^2} &=& (2.90 +1.24 -1.19 ,\, 55 +16 -15 ,\, 2\,900. \pm 90. , \;\,
300\,000.\leq m_{d_4}  \leq 1\, 200 \,000.)\,.\nonumber\\ 
\end{eqnarray} 
The matrix elements of the $4 \times 4$ quark mixing matrix are determined in the numerical 
procedure, which searches for the best fit of the two quarks mass matrices free parameters, 
presented in Eq.~(\ref{M0}), to the experimental data, by taking into account the experimental 
inaccuracy and the unitarity, in this paper orthogonality, of the $4 \times 4$ mixing matrix, 
ensuring as much as possible, the best fit as defined in Eq.~(\ref{sigma}).

Let us notice that the new experimental data for the quark mixing matrix differ from the old 
ones the most in the two diagonal matrix elements, $V_{cs}=V_{u_2 d_2}$ and $V_{tb}$ 
$= V_{u_3 d_3}$, appearing in the new data with smaller inaccuracy. The differences among
the old and the new values of $V_{us}=V_{u_1 d_2}$, $V_{ub}= V_{u_1 d_3}$, 
$V_{cd}=V_{u_2 d_1}$, $V_{cb}= V_{u_2 d_3}$ and $V_{ts}= V_{u_3 d_2}$, are  smaller 
(some of them even with not better accuracy than in the old data) than
in the case of the diagonal ones, 
while the rest two, $V_{ud}= V_{u_1 d_1}$ and $V_{td}= V_{u_3 d_1}$, were not measured 
more accurately. The values for the quark masses  are taken in both cases within the measured 
inaccuracy. 

The corresponding fourth family mixing matrix elements ($|V_{u_i d_4}|$ and $|V_{u_4 d_j}|$)
are in both cases determined from the unitarity condition for the $4 \times 4$ mixing matrix
through the fitting procedure,  explained in Sect.~\ref{procedure}, as also all the other matrix
elements of the mixing matrix are.

Using first the old experimental data~\cite{dataold} we predict the direction in which new more
accuratelymeasured matrix elements should move and then we check if this is happening with the
new experimental data~\cite{datanew}.

Then we use the new experimental data, repeat the variational procedure and look for what are 
our new  results predicting.

The results are presented in the next Subsect.~\ref{quarks}.

\subsection{The mass matrices for quarks, their masses, the mixing matrix and predictions}
\label{quarks}

As already written, the mixing matrix elements for quarks, forming in the {\it spin-charge-family} 
$3 \times 3$ submatrix of the $4 \times 4$ unitary matrix, are not yet measured accurate 
enough to allow us a trustworthy prediction of the fourth family quark masses. Correspondingly
we perform all the calculation for the chosen set of the fourth family masses and study how does
the accuracy of the fitting procedure, explained in Sect.~\ref{procedure}, depend on the fourth 
family masses.

Still we can make predictions: {\bf i.} We make a very rough estimation of the fourth family 
masses. {\bf ii.} We evaluate the matrix elements $V_{u_{i} d_{4}}$ and $V_{u_{4} d_{j}}$, 
$(i,j) \in (1,2,3)$ for chosen fourth family quark masses and check their dependence on all 
the quark masses (with the fourth family included). {\bf iii.} We predict how will the matrix elements
 $V_{u_{i} d_{j}}$ of the $3 \times 3$ submatrix of the $4 \times 4$ mixing matrix change in
next more accurate measurements, under the assumption that the {\it spin-charge-family} theory
is the right next step beyond the {\it standard model}.

We test the extent to which our results have some experimental support by performing 
calculations with the two kinds of the experimental values for the quark mixing matrix: The older 
ones presented in Eq.~(\ref{vudold}) and the newer ones presented in Eq.~(\ref{vudnew}). 
We use both data with the same set of the fourth family quark masses. 

Below we present results for the two choices of masses: for $m_{u_4} = 700$ GeV $=m_{d_4}$
and for $m_{u_4} = 1\,200$ GeV $=m_{d_4}$, obtained when fitting twice $6$ free parameters 
of the mass matrices to twice three measured masses of quarks and to their measured mixing 
matrix, first for the old  data~\cite{dataold} and then for the new data~\cite{datanew}, taking into 
account the experimental accuracy, although we test the whole interval for the fourth family
 masses, from $300$ GeV to $1\,700$ GeV. We present the two choices as an  illustration. 

Looking for the results from the fitting procedure when using the old (\cite{dataold}) experimental 
data for the quark mixing matrix, we predict, from the calculated $3 \times 3$ submatrix, the
expected changes in the new data (\cite{datanew}) and comment these predictions. Then we 
repeat calculations with the new data (\cite{datanew}) and predict, how will the matrix elements 
change in next more accurate measurements.

We keep the symmetry (Eq.~(\ref{M0})) of mass matrices exact. 
\begin{itemize}
\item  $\,\,\,$ {\bf I.}
 We present first the results of our calculations when using the old data~\cite{dataold}
(Eqs.~(\ref{mumd}, \ref{vudold})), and make a choice for the fourth family quark masses: 
first $m_{u_4} = m_{d_4}=700$ GeV 
and then $m_{u_4} = m_{d_4}=1\,200$ GeV. We present for the best fit (Eq.~(\ref{sigma}))
to the old experimental data the corresponding mass matrices, their diagonal values, the mixing
matrix and the deviations of the calculated matrix elements from the measured ones
 (Eq.~(\ref{sigma})). 
\begin{enumerate} 
\item 
We choose first $m_{u_4} = 700$ GeV and $m_{d_4}=700$ GeV and obtain through  the variational procedure (Sect.~\ref{procedure}) the two mass matrices. 
\begin{equation}
\label{mmudold1}
 M^{u} = \begin{pmatrix} 
227 623. & 131 877. & 132 239. & 217 653.\\ 
131 877. & 222 116. & 217 653. & 132 239.\\ 
132 239. & 217 653. & 214 195. & 131 877.\\ 
217 653. & 132 239. & 131 877. & 208 687. 
\end{pmatrix}\,,
M^{d} = \begin{pmatrix} 
175 797. & 174 263. & 174 288. & 175 710.\\ 
174 263. & 175 666. & 175 710. & 174 288.\\ 
174 288. & 175 710. & 175 813. & 174 263.\\ 
175 710. & 174 288. & 174 263. & 175 682.
\end{pmatrix}\,,
\end{equation} 
which define the mixing matrix 
\begin{equation}
\label{vudold1}
 V_{ud}= \begin{pmatrix}
-0.97423 &  0.22531 & -0.00299 & 0.01021\\  
 0.22526 &  0.97338 & -0.04238 & 0.00160\\ 
-0.00663 & -0.04197 & -0.99910 & -0.00040\\ 
 0.00959 & -0.00388 & -0.00031 &  0.99995 
\end{pmatrix}\,, 
\end{equation} 
and the corresponding absolute values for the deviations from the average experimental values 
(Eq.~(\ref{sigma})) 
\begin{equation} 
\label{vudold1dev} 
\delta V_{ud}= \begin{pmatrix} 
   0.091 & 0.117 & 2.339\\ 
   0.431 & 1.418 & 1.348\\ 
   2.951 & 0.358 & 1.559 
\end{pmatrix}\,. 
\end{equation}
The corresponding total absolute average deviation, defined in Eq.~(\ref{sigma}), is $4.55785$.\\ 
The diagonal values of the two mass matrices from Eq.~(\ref{mmudold1}) determine quark 
masses 
\begin{eqnarray} 
\label{mdudold1}
{\bf M}^{u}_{d}/{\rm MeV/c^2} &=& (1.3, 620.0, 172\,000. , 700\,000.)\,,\nonumber\\ 
{\bf M}^{d}_{d}/{\rm MeV/c^2} &=& (2.88508 , 55.024 , 2\,899.99 , 700\,000.)\,. \end{eqnarray}
\item
We next choose $m_{u_4} = 1\,200$ GeV and $m_{d_4}=1\,200$ Ge and fit the 
parameters of the two quark mass matrices (Eq.~(\ref{M0})) to the old experimental 
data~\cite{dataold}. We obtain the following two mass matrices 
\begin{equation}
 \label{mmudold2} 
M^{u} = \begin{pmatrix}
 351 916. & 256 894. & 257 204. & 342 714.\\
 256 894. & 344 411. & 342 714. & 257 204.\\ 
 257 204. & 342 714. & 341 900. & 256 894.\\ 
 342 714. & 257 204. & 256 894. & 334 395.
\end{pmatrix}\,,
M^{d} = \begin{pmatrix} 
 300 783. & 299 263. & 299 288. & 300 709.\\ 
 299 263. & 300 623. & 300 709. & 299 288.\\ 
 299 288. & 300 709. & 300 856. & 299 263.\\ 
 300 709. & 299 288. & 299 263. & 300 696. 
\end{pmatrix}\,, 
\end{equation}
and the mixing matrix 
\begin{equation}
\label{vudold2} 
V_{ud}= \begin{pmatrix}
 -0.97425 &  0.22536 & -0.00301 &  0.00474\\
  0.22534 &  0.97336 & -0.04239 &  0.00212\\
 -0.00663 & -0.04198 & -0.99910 & -0.00021\\
  0.00414 & -0.00315 & -0.00011 &  0.99999 
\end{pmatrix}\,.
\end{equation} 
The corresponding values for the deviations from the average experimental value of the matrix elements of the $3 \times 3$ submatrix are 
\begin{equation}
\label{vudold2dev} 
\delta V_{ud}= \begin{pmatrix}
  0.003 & 0.226 & 2.335\\
  0.424 & 1.419 & 1.357\\ 
  2.949 & 0.355 & 1.559 
\end{pmatrix}\,. 
\end{equation} 
The corresponding total average deviation (Eq.~(\ref{sigma})) is now $4.55955$.\\
The diagonal values of the two mass matrices from Eq.~(\ref{mmudold2}) determine quark 
masses 
\begin{eqnarray} 
\label{mdudold2}
{\bf M}^{u}_{d}/{\rm MeV/c^2} &=& (1.3 , 620.0 , 172\,000. , 1\,200\,000.)\,,\nonumber\\ 
{\bf M}^{d}_{d}/{\rm MeV/c^2} &=& (2.9 , 55.0 , 2\,900.0 , 1\,200\,000.)\,. 
\end{eqnarray} 
\end{enumerate}
One notices, that while the matrix elements of mass matrices of the $u$ and the $d$ quark
change for a factor of $\approx 1.5$ when changing the fourth family masses from $700$ GeV
to $1\,200$ GeV, becoming more "democratic" (that is the matrix elements become more and
more equal), the mixing matrix elements of the $3\times 3$ submatrix change  very little 
(Eqs.~(\ref{vudold1}, \ref{vudold2})).  Let us add that the  calculations, repeated with within
the experimentally allowed intervals of twice three quark masses do not influence the results
noticeably.

Let us now analyze the  results obtained with the old data~\cite{dataold}, Eq.~(\ref{vudold}), 
for the two choices of the fourth family masses: $700$ GeV and $1\,200$ GeV. In both cases
the rest two times three masses (Eqs.~(\ref{mdudold1}, \ref{mdudold2})) are the same and
 in agreement with the experimental data (Eq.~(\ref{mumd})). Variation of the three measured 
masses, in particular of the lowest $5$ ($m_{u}, m_{c}, m_{d}, m_{s}, m_{b}$),  within the
 experimentally declared intervals does not influence the mixing matrix noticeable. We shall 
demonstrate this with the new experimental results, when changing $m_{t}$ for $3 \times 760$ MeV. 
 
Let us, therefore, compare the two calculated mixing matrices, Eqs.~(\ref{vudold1}, 
\ref{vudold2}), with the old measured ones, Eq.~(\ref{vudold}). In Eq.~(\ref{vudoldexp1}) these 
three kinds of mixing matrix elements are denoted by  ($exp_o$, $old_1$ and $old_2$) for the 
old experimental  data, and the two calculated ones for the choice for the fourth family masses: 
$700$ GeV and $1\,200$ GeV, respectively.
    \begin{equation}
     \label{vudoldexp1}
     |V_{(ud)_{old}}|= \begin{pmatrix}
    exp_o &   0.97425 \pm 0.00022    &  0.2252 \pm 0.0009    &  0.00415 \pm 0.00049     \\
    \hline
    old_1 &   0.97423                &  0.22531              &  0.00299\\
    old_2 &   0.97425                &  0.22536              &  0.00301\\
    \hline 
    exp_o &   0.230   \pm 0.011      &  1.006  \pm 0.023     &  0.0409  \pm 0.0011     \\
   \hline
   old_1  &   0.22526                &  0.97338              &  0.04238 \\
   old_2  &   0.22534                &  0.97336              &  0.04239\\ 
   \hline
   exp_o  &   0.0084  \pm 0.0006     &  0.0429 \pm 0.0026    &  0.89    \pm 0.07      \\ 
  \hline
   old_1  &   0.00663                &  0.04197              &  0.99910\\
   old_2  &   0.00663                &  0.04198              &  0.99910   
    \end{pmatrix}\,.
    \end{equation}
 Comparing the two calculated mixing matrix elements ($old_1$ with $m_{u_{4}}=$ 
 $m_{d_{4}}= 700$ GeV and with $m_{u_{4}}=$  $m_{d_{4}}=1\,200$ GeV) with the 
measured ones,  
Eq.~(\ref{vudold}), we see:\\
{\bf a.} The calculated matrix elements of the $3\times 3$ submatrix of the $4\times 4$ mixing 
matrix do not depend much on the masses of the fourth family members. As expected, they
rise with the raising fourth family masses, but slightly, while the fourth family matrix elements, 
$V_{u_i d_4}$ and $V_{u_4 d_i}$, decrease, with the exception of one of them which even
becomes larger, $V_{u_2 d_4}$.  The accuracy with which we fitted the experimental data, 
(Eqs.~(\ref{vudold1dev}, \ref{vudold1dev}), smaller numbers mean better fitting, the number 
$2$ means twice worse, while $0.1$ means $10$ times better fitting than it is the experimental 
inaccuracy) are not in average better (some are better, the others are worse).  \\
{\bf b.}  
{\it  Comparing the measured and the
calculated matrix elements of the $3\times 3$ submatrix  we make prediction for next (now 
already known) measurements}~\cite{datanew}:\\
{\bf b.i.} The matrix element $V_{u_{1} d_{1}}$ ($V_{ud}$) would very slightly decrease or stay 
unchanged,
$V_{u_{1} d_{2}}$ ($V_{u s}$) will rise a little bit, and $V_{u_{2} d_{3}}$ ($V_{c b}$) 
as well as $V_{u_{3} d_{3}}$  ($V_{t b}$) will rise more. \\
{\bf b.ii.}
The matrix elements $V_{u_{1} d_{3}}$  ($V_{u b}$), $V_{u_{2} d_{1}}$  ($V_{c d}$),
 $V_{u_{2} d_{2}}$  ($V_{c s}$), $V_{u_{3} d_{1}}$  ($V_{t d}$) and 
$V_{u_{3} d_{2}}$  ($V_{t s}$) will lower.\\
{\it Checking the new experimental values}, Eq,(\ref{vudold}),{\it  with these predictions one sees 
that the prediction is in agreement with the new experimental data}, Eq.~(\ref{vudnew}).

 \item  $\,\,\,$ {\bf II.} 
Let us repeat the calculations with new experimental data~\cite{datanew} (Eqs.~(\ref{mumd}, 
 \ref{vudnew})) to see how will the new data influence the mass matrices and the mixing matrix 
elements. Again we use the same two choices for the fourth family masses  (they have quite
 different values and are enough illustrative, since any other two equally separated choices would
lead to the similar recognitions), $m_{u_4} $  $= m_{d_4}= 700$ GeV and $m_{u_4} = $ 
$m_{d_4}=1\,200$ GeV.  
 
Results of the calculations with the new experimental data (Eqs.~(\ref{mumd}, \ref{vudnew})) 
 show smaller common deviations for the sums of all the average values of  the nine matrix 
elements of the $3 \times 3$ submatrix (Eq.~(\ref{sigma})) -   
$4.0715$ and $4.0724$ - than those with the old data - $4.5579$ and $4.5596$.
%
 \begin{enumerate}
\item For $m_{u_4} = 700$ GeV and $m_{d_4}=700$ GeV we get for the mass matrices  
 \begin{equation}
 \label{mmudnew1}
  M^{u} = \begin{pmatrix}
  226 521. & 131 887. & 132 192. & 217 715.\\
  131 887. & 219 347. & 217 715. & 132 192.\\
  132 192. & 217 715. & 216 964. & 131 887.\\
  217 715. & 132 192. & 131 887. & 209 790.
 \end{pmatrix}\,,
  M^{d} = \begin{pmatrix}
  175 776. & 174 263. & 174 288. & 175 709.\\
  174 263. & 175 622. & 175 709. & 174 288.\\
  174 288. & 175 709. & 175 857. & 174 263.\\
  175 709. & 174 288. & 174 263. & 175 703.
\end{pmatrix}\,,  
\end{equation}
for the mixing matrix
\begin{equation}
 \label{vudnew1}
 V_{ud}= \begin{pmatrix}
 -0.97423 &  0.22539 & -0.00299 &  0.00776\\
  0.22534 &  0.97335 & -0.04245 &  0.00349\\
 -0.00667 & -0.04203 & -0.99909 & -0.00038\\
  0.00677 & -0.00517 & -0.00020  &  0.99996
 \end{pmatrix}\,.
\end{equation}
The corresponding values (Eq.~(\ref{sigma})) for the deviations from the average experimental 
values are
\begin{equation}
 \label{vudnew1dev}
 \delta V_{ud}= \begin{pmatrix}
 & 0.074 &  0.109 & 2.336\\
 & 0.043 &  0.791 & 1.042\\
 & 2.891 &  0.753 & 0.685 
 \end{pmatrix}\,,
\end{equation}
and the corresponding total absolute average deviation Eq.~(\ref{sigma}) is  $4.07154$.\\

The two mass matrices correspond to the diagonal masses 
\begin{eqnarray}
\label{mdudnew1}
 {\bf M}^{u}_{d}/{\rm MeV/c^2} &=& (1.3 , 620.0 , 172\,000. ,   700\,000.)\,,\nonumber\\ 
 {\bf M}^{d}_{d}/{\rm MeV/c^2} &=& (2.9 , 55.0 ,  2\,900.0 ,   700\,000.)\,. 
 \end{eqnarray}
 %

 \item For $m_{u_4} = 1\,200$ GeV and $m_{d_4}=1\,200$ GeV we obtain  the mass matrices
  \begin{equation}
  \label{mmudnew2}
   M^{u} = \begin{pmatrix} 
 354 761. & 256 877. & 257 353. & 342 539.\\
 256 877. & 350 107. & 342 539. & 257 353.\\
 257 353. & 342 539. & 336 204. & 256 877.\\
 342 539. & 257 353. & 256 877. & 331 550.
  \end{pmatrix}\,,
   M^{d} = \begin{pmatrix}
 300 835. & 299 263. & 299 288. & 300 710.\\
 299 263. & 300 714. & 300 710. & 299 288.\\
 299 288. & 300 710. & 300 765. & 299 263.\\
 300 710. & 299 288. & 299 263. & 300 644.
 \end{pmatrix}\,, 
 \end{equation}
 and the mixing matrix 
 \begin{equation}
  \label{vudnew2}
  V_{ud}= \begin{pmatrix}
   0.97423  &  0.22538 & 0.00299  &  0.00793\\
  -0.22514  &  0.97336 & 0.04248  & -0.00002\\
   0.00667  & -0.04206 & 0.99909  & -0.00024\\
  -0.00773  & -0.00178 & 0.00022  &  0.99997
  \end{pmatrix}\,.
\end{equation} 
 The corresponding values for the deviations from the average experimental value for each matrix
 element are
 \begin{equation}
  \label{vudnew2dev}
  \delta V_{ud}= \begin{pmatrix}
  0.070  & 0.097 & 2.329\\
  0.038 & 0.790  & 1.061\\
  2.889 & 0.762 & 0.685
  \end{pmatrix}\,. %
 \end{equation}
 The corresponding total average deviation Eq.~(\ref{sigma}) is $4.0724$.\\
 
 The two mass matrices correspond to the diagonal masses 
 \begin{eqnarray}
 \label{mdudnew2}
  {\bf M}^{u}_{d}/{\rm MeV/c^2} &=& (1.3 , 620.0 , 172\,000. ,   1\,200\,000.)\,,\nonumber\\ 
  {\bf M}^{d}_{d}/{\rm MeV/c^2} &=& (2.88508 , 55.024 ,  2\,899.99 ,   1\,200\,000.)\,. 
  \end{eqnarray}
 \end{enumerate}

\item  $\,\,\,$ {\bf III.} Let us check the sensitivity of our results with respect to changes of
the measured quark masses within experimental inaccuracy. The influence of the changes of
the lowest measured masses ($m_{u}, m_{d}, m_{s}, m_{c}, m_{b}$) within experimental 
inaccuracy is not noticeable, the only one which can change the results is the top mass. 
We therefore present the two calculated mixing matrices with  $m_{t} = (172 \pm 
3\times 0.760) $ GeV fitted to the new experimental data~\cite{datanew}.
\begin{enumerate}
%
\item First we do calculations for  $m_{u_4} = $ $m_{d_4}=700$ GeV and $m_{u_3}=m_t$ $=(172 + 3\times 0.760)$ GeV with the new experimental data~\cite{datanew}. 
Below are the calculated mixing matrix
\begin{equation}
  \label{vudnew760}
  V_{ud}= \begin{pmatrix}
  -0.97424  &  0.22537 & -0.00299  &  0.00771\\
   0.22533  &  0.97335 & -0.04246  &  0.00360\\
  -0.00666  & -0.04206 & -0.99909  & -0.00024\\
  -0.00670  & -0.00526 & -0.00020  &  0.99996
  \end{pmatrix}\,,
\end{equation} 
 and the corresponding values for the deviations from the average experimental value for 
 each matrix element 
 \begin{equation}
  \label{vudnewdev760}
  \delta V_{ud}= \begin{pmatrix}
  0.057  & 0.091 & 2.330\\
  0.041 & 0.791  & 1.046\\
  2.895 & 0.755 & 0.685
  \end{pmatrix}\,. 
 \end{equation}
 The corresponding total average deviation Eq.~(\ref{sigma}) is $4.07154$.\\
 
 The two mass matrices correspond to the diagonal masses 
 \begin{eqnarray}
 \label{mdudnew760}
  {\bf M}^{u}_{d}/{\rm MeV/c^2} &=& (1.3 , 620.0 , 174\,260. ,  700\,000.)\,,\nonumber\\ 
  {\bf M}^{d}_{d}/{\rm MeV/c^2} &=& (2.88508 , 55.024 ,  2\,899.99 ,  700\,000.)\,. 
  \end{eqnarray}
  %

\item We repeat the  calculations for  $m_{u_4} = $ $m_{d_4}=1\,200$ GeV and 
$m_{u_3}=m_t$ $=(172 - 3\times 0.760)$ GeV with the new experimental 
data~\cite{datanew}. 
Below are the calculated mixing matrix
\begin{equation}
  \label{vudnew1200}
  V_{ud}= \begin{pmatrix}
   0.97425  &  0.22542 & 0.00299  &  0.00466\\
  -0.22535  &  0.97335 & 0.04248  & -0.00216\\
   0.00667  & -0.04205 & 0.99909  & -0.00021\\
  -0.00405  & -0.00526 &-0.00010  &  0.99999
  \end{pmatrix}\,,
\end{equation} 
and the corresponding values for the deviations from the average experimental value for
each matrix element 
 \begin{equation}
  \label{vudnewdev1200}
  \delta V_{ud}= \begin{pmatrix}
  0.017  & 0.146 & 2.336\\
  0.044 & 0.791  & 1.058\\
  2.883 & 0.761 & 0.685
  \end{pmatrix}\,. 
 \end{equation}
 The corresponding total average deviation Eq.~(\ref{sigma}) is $4.07281$.\\
 
 The two mass matrices correspond to the diagonal masses 
 \begin{eqnarray}
 \label{mdudnew1200}
  {\bf M}^{u}_{d}/{\rm MeV/c^2} &=& (1.3 , 620.0 , 169\,740. ,   1\,200\,000.)\,,\nonumber\\ 
  {\bf M}^{d}_{d}/{\rm MeV/c^2} &=& (2.88508 , 55.024 ,  2\,899.99 ,   1\,200\,000.)\,. 
  \end{eqnarray}
\end{enumerate}

One notices, that in the case that the new experimental data~\cite{datanew} are used in the 
calculation, the matrix elements of mass matrices of the $u$-quarks differ less from those of the
$d$-quarks than in the case when the old experimental data are used. The new experimental data
lead to even more "democratic" mass matrices than the old ones, in particular is this the case for 
$m_{u_4}= m_{d_4} =1\,200$ GeV.\\ 
The mixing matrix elements of the $3\times 3$ submatrix calculated with the new data (those 
obtained with $m_{u_4}= m_{d_4} = 700$ GeV are again close to those obtained with  
$m_{u_4}= m_{d_4} =1\,200$ GeV (Eqs.~(\ref{vudnew1}, \ref{vudnew2}))) agree better 
with the newer~\cite{datanew} than with the older~\cite{dataold} experimental values, as we 
predicted (results  are presented in Eq.~(\ref{vudoldexp1}) and predictions made in {\bf b.} 
below this equation) and already recognized.  

Comparing the calculated mixing matrix elements obtained with $m_{u_4}=m_{d_4}= 700 $
GeV and either with $m_t =172$ GeV (Eq.~(\ref{vudnew1}))  or with $m_t =$ $(172+3\times0.76)$
GeV (Eq.~(\ref{vudnew760})), as well as by comparing the calculated mixing matrix elements 
obtained with $m_{u_4}=m_{d_4}= 1\,200 $ GeV and  with either 
$m_t =172$ GeV (Eq.~(\ref{vudnew1})) or with $m_t =(172-3\times 0.76)$ GeV  
(Eq.~(\ref{vudnew760})) one sees that the $3\times 3$ submatrix changes in both cases
very little, changes are almost negligible. The matrix elements $V_{u_{i} d_4}$ and
 $V_{u_{4} d_i}$ are, as expected, much stronger influenced.
 \end{itemize}

We present in Eq.~(\ref{vudoldnewexp}) the matrix elements of the $4\times 4$ mixing matrix 
for quarks obtained when the $4\times 4$ mass matrices respect the symmetry of 
Eq.~(\ref{M0}) while we fit the parameters of the mass matrices to the old ($exp_o$)  and the
new ($exp_n$) experimental data. In both cases we present results for the choices of the fourth 
family quark masses:  $m_{u_4}= m_{d_4}=700$ GeV  ($old_{1}, new_{1}$) and 
$m_{u_4}= m_{d_4}=$ $1\,200$ GeV ($old_{2}, new_{2}$). In parentheses, $(\;)$ and $[\;\,]$,
the  changes of the matrix elements are presented, which are due to the changes of the top mass
within the experimental inaccuracies: with the $m_{t} =$ $(172 + 3\times 0.76)$ GeV and $m_{t} =$ 
$(172 - 3\times 0.76)$, respectively  (if there is one number in parentheses only the last number 
is different, if there are two or more numbers in parentheses the last two or more numbers 
are different, if there is no parentheses no numbers are different). 
\begin{equation}
\label{vudoldnewexp}
      |V_{(ud)}|= \begin{pmatrix}
     exp_o &   0.97425 \pm 0.00022    &  0.2252 \pm 0.0009   &  0.00415 \pm 0.00049 &    \\
     exp_n  &    0.97425 \pm 0.00022    &  0.2253 \pm 0.0008 &  0.00413 \pm 0.00049&   \\
     \hline
     old_1 &       0.97423                &  0.22531              &  0.00299  & 0.01021\\
     old_2 &       0.97425                &  0.22536              &  0.00301  & 0.00474\\
     new_1  &    0.97423(4)            &  0.22539(7)          &  0.00299  & 0.00776(1)\\  
     new_2  &    0.97423[5]            &  0.22538[42]        &  0.00299  & 0.00793[466]\\ 
     \hline 
     exp_o &   0.230   \pm 0.011      &  1.006  \pm 0.023     &  0.0409  \pm 0.0011&     \\
     exp_n  &  0.225   \pm 0.008      &  0.986  \pm 0.016     &  0.0411  \pm 0.0013&   \\
    \hline
    old_1  &   0.22526            &  0.97338              &  0.04238    & 0.00160 \\
    old_2  &   0.22534            &  0.97336              &  0.04239    & 0.00212 \\
    new_1  &  0.22534(3)       &  0.97335              &  0.04245(6) & 0.00349(60) \\  
    new_2  &  0.22531[5]       &  0.97336[5]          &  0.04248     & 0.00002[216] \\ 
    \hline
    exp_o  &   0.0084  \pm 0.0006     &  0.0429 \pm 0.0026    &  0.89    \pm 0.07&      \\ 
    exp_n  &  0.0084  \pm 0.0006     &  0.0400 \pm 0.0027    &  1.021   \pm 0.032&     \\
   \hline
    old_1  &   0.00663                &  0.04197              &  0.99910   &0.00040 \\
    old_2  &   0.00663                &  0.04198              &  0.99910   &0.00021\\
    new_1  &  0.00667(6)            &  0.04203(4)         &  0.99909   & 0.00038\\  
    new_2  &  0.00667                &  0.04206[5]         &  0.99909   & 0.00024[21] \\
   \hline
   old_1    & 0.00959                  & 0.00388             & 0.00031    & 0.99995\\
   old_2    & 0.00414                   & 0.00315             & 0.00011    &  0.99999\\
   new_1   & 0.00677(60)             & 0.00517(26)        & 0.00020    & 0.99996\\
   new_2   & 0.00773                  & 0.00178             & 0.00022    & 0.99997[9]
     \end{pmatrix}\,.
     \end{equation}

Comparing the calculated mixing matrix elements for quarks, obtained  by taking into account by
the {\it spin-charge-family} theory suggested symmetry of  mass matrices  (Eq.~(\ref{M0})) with
the measured ones, to which the calculated values are fitted, we notice:\\
{\bf A.} The matrix elements of the $3\times 3$ submatrix of the $4\times 4$ mixing matrix 
depend very little on the masses of the fourth family members. As expected, they do rise
with the raising fourth family masses, but  very slightly, while the fourth family matrix elements, 
$V_{u_i d_4}$ and $V_{u_4 d_i}$, $i=(1,2,3)$, decrease (and correspondingly  
$V_{u_4 d_4}$ increase) in general, as expected, with the  fourth family masses. But not
all of them, the results with the new experimental data show that $V_{u_1 d_4}$, 
 $V_{u_4 d_1}$  as well as  $V_{u_4 d_3}$ even rise  with the fourth family masses. 
The fourth family matrix elements are  sensitive also to the top quark mass.  
The accuracy with which we fitted the experimental data, [Eqs.~(\ref{vudnew1dev}, 
\ref{vudnew2dev})] is better with the new experimental data and approximately the same 
for the two choices of the fourth family masses. The above (in item {\bf b.}) predicted 
changes of mixing matrix elements were realized with the more accurate experimental
data~\cite{datanew}. \\
{\bf B.} Changes of the lowest five quark masses within the experimental inaccuracy do not
 influence the results noticeably (and are correspondingly not presented among the
calculated values). The presented results for the choice of the $m_{t}$ mass
 within $ 3 \sigma$ (Eqs.~(\ref{vudnew760}, \ref{vudnewdev760}, \ref{vudnew1200}, 
\ref{vudnewdev1200})) show that $3 \times 3$ submatrix elements almost do not change 
within the experimental accuracy changed top mass. The elements $V_{u_i d_4}$ and 
$V_{u_4 d_i }$ are sensitive to the experimental accuracy of the top mass.\\
{\bf C.}  {\it  Comparing the measured and the calculated matrix elements, of the 
$3\times 3$ submatrix of the $4\times 4$ mixing matrix we are making predictions for next 
more accurate measurements}:\\
%
{\bf C.i.} $\;\,\,$ The matrix element $V_{u_{1} d_{1}}$ ($V_{ud}$) will stay the same or
will very slightly decrease,  $V_{u_{1} d_{2}}$ ($V_{u s}$) will still very slightly increase, 
$V_{u_{2} d_{1}}$ ($V_{c d}$) will (after decreasing in $exp_{n}$~\cite{datanew}) very
slightly increase towards $exp_{o}$, 
$V_{u_{2} d_{3}}$ ($V_{c b}$)  will still increase, and $V_{u_{3} d_{2}}$  ($V_{t s}$)
will (after decreasing in $exp_{n}$) slightly rise again (in the new experimental 
data~\cite{datanew}, with worth accuracy, it is now too low). \\
{\bf C.ii.} $\;\,\,$ The matrix elements $V_{u_{1} d_{3}}$  ($V_{u b}$) and  
$V_{u_{2} d_{2}}$  ($V_{c s}$) will still lower, $V_{u_{3} d_{1}}$  ($V_{t d}$) should lower and  
$V_{u_{3} d_{3}}$  ($V_{t b}$) must again lower.\\
{\bf C.iii.} $\;\,\,$ The fourth family masses change the mass matrices considerably, while
 their influence on the $3 \times 3$ submatrix of the $4 \times 4$ mixing matrix is quite 
weak. 
Accordingly there is very difficult to predict the fourth family masses from the today 
experimental data. We only can say, taking into account all the experimental evidences,
that they might be around $1$ TeV or above.\\

 We can conclude: Requiring that the experimental data respect the symmetry of the mass
 matrices  (Eq.~(\ref{M0})) (suggested by the {\it spin-charge-family} theory) the prediction
 can be made  for the changes of the matrix elements of the $3 \times 3$ submatrix in future 
experiments. {\it The masses  of the fourth family members are more difficult to be predicted, 
since they are very sensitive} to the accuracy of the experimental  data of  the quark
masses (only the top mass counts) and in particular to {\it the accuracy of the experimental data 
of the quark mixing matrix}, which is now too low. 
{\it  For the known fourth family masses the fourth family matrix elements of the mixing matrix
 can be predicted}.  The fourth family matrix elements of the mixing matrix are even not very 
sensitive to the choice of the fourth family masses.

\section{Discussions and conclusions}
\label{discussions}

One of the most important open questions in the elementary particle physics is: Where do 
the families originate? Explaining the origin of families would answer the question about the 
number of families which are possibly observable at the low energy regime, about the origin 
of the scalar field(s) and Yukawa couplings and would also explain differences in the fermions 
properties - the differences in masses and mixing matrices among family members --
quarks and leptons.  

The {\it spin-charge-family} theory is offering a possible answer to  the questions about the
origin of families, of the scalar fields and the Yukawa couplings, as well as to several additional  
open questions of the {\it standard model} and also beyond 
it~\cite{norma2014MatterAntimatter}. This theory predicts that there are four rather
than so far observed  three coupled families. The mass matrices of all the family members 
(quarks and leptons) demonstrate in the massless basis the $U(1)\times SU(2)\times SU(2)$
(each of the two $SU(2)$ is a subgroup, one of $SO(1,3)$ and the other of $SO(4)$) 
symmetry,  Eq.~(\ref{M0}). 

Any accurate $3\times 3$ submatrix of the $4 \times 4$ unitary matrix determines the 
$4 \times 4$ matrix uniquely. Since neither the quark and (in particular) nor the lepton 
$3\times 3$ mixing matrix are measured accurately enough to be able to determine three
complex phases of the $4 \times 4$ mixing matrix, we assume (what also simplifies the 
numerical procedure)  that the mass matrices are symmetric and real and correspondingly
the mixing matrices are orthogonal. We fitted the $6$ free parameters of each family member
mass matrix, Eq.~(\ref{M0}),  to twice three measured masses ($6$) of each pair of either 
quarks or leptons and to the $6$ (from the experimental data extracted) parameters of the 
corresponding $4 \times 4$ mixing matrix.

We are presenting the results for quarks only. The accuracy of the experimental data for 
leptons are not yet accurate enough that would allow us meaningful predictions. 

The numerical procedure, explained in this paper, to fit free parameters to the experimental 
data within the experimental inaccuracy of masses and in particular of the mixing matrices 
is very tough. 
It turned out that the experimental inaccuracies are too large to tell trustworthy mass
intervals for the quarks masses of the fourth family members. Taking into account our 
calculations fitting
the experimental data (and the meson decays evaluations in literature, as well as our own)
we very roughly  estimate that the fourth family quarks masses might be around $1$ TeV 
or above.

Since the matrix elements of the $3 \times 3$ submatrix of the $4 \times 4$ mixing 
matrix depend very weakly on the fourth family masses, the calculated mixing matrix
offer the prediction to what values will more accurate measurements move the present 
experimental data and also the fourth family mixing matrix elements in dependence of the
fourth family masses, Eq.~(\ref{vudoldnewexp}). The predictions are presented in 
Subsect.~\ref{quarks}, in particular in Eq.~(\ref{vudoldnewexp}), 
and in comments below this equation. 

We expect - detailed values can be found in Eq.~(\ref{vudoldnewexp}) - that more accurate 
experiments will bring in comparison with the data~\cite{datanew}: a slightly smaller values 
for $V_{u d}$;  $V_{u b}$ and $V_{c s}$ will still lower, $V_{t d}$ will lower, $V_{t b}$  will
now lower; $V_{u s}$ will still slightly rise, $V_{c d}$ will now slightly rise,  $V_{c b}$ will 
still rise and $V_{t s}$ will now (after decreasing) increase. 

We checked our ability to make predictions for the real experiment by first performing 
calculations  with the old experimental data~\cite{dataold} and test the predictions on 
the new experimental data~\cite{datanew}. The results, presented in 
Eq.~(\ref{vudoldexp1}) and commented below  Eq.~(\ref{vudoldexp1}) ({\bf b.}), 
manifest that our predictions are in agreement with  the new experimental 
data~\cite{datanew}. 

The fourth family mixing matrix elements  depend, as expected, quite strongly on the fourth family
masses. With the increasing fourth family masses they decrease, but not all (see 
Eq.~(\ref{vudoldnewexp})), $V_{u_1 d_4}$, $V_{u_4 d_1}$, and $V_{u_4 d_3}$ even 
rise with the rising four family masses.
For chosen  (quite large intervasl of the) masses of the fourth family members  are their matrix
 elements quite accurately 
predicted (Eqs.~(\ref{vudnew1}, \ref{vudnew2})).

Mass matrices are quite close to the "democratic" ones not only for leptons (which we not 
present in this paper) but also for quarks, Eqs.~(\ref{mmudnew1}, \ref{mmudnew2}). 
With the growing fourth family masses the "democracy" in matrix elements 
grow (Eqs.~(\ref{mmudnew1}, \ref{mmudnew1})), as expected. 

The complex mass matrices would lead to unitary and not to orthogonal mixing matrices.
The more accurate experimental data for quarks would allow us to extract also the phases of
the unitary mixing matrices, changing as well our predictions. In particular they will allow us
to predict the fourth family masses much more accurately.

More accurately experimental data for leptons would allow us to make predictions also 
for leptons and to manifest that the {\it spin-charge-family} theory is explaining the properties of 
of all the family members.

\appendix
\section{A brief presentation of the {\it spin-charge-family} theory}
 \label{scft}

We present in this section a very brief introduction into the {\it spin-charge family} 
theory~\cite{NBled2013,NBled2012,norma92,norma93,norma94,pikanorma,JMP,norma95,%
gmdn07,gn,gn2013,NPLB,N2014scalarprop,norma2014MatterAntimatter} to explain to
readers the origin and properties of the families and correspondingly of the Higgs's scalar,
with the weak and the hyper charge equal to ($\pm \frac{1}{2}$, $\mp \frac{1}{2}$), 
respectively, and the Yukawa couplings. This theory predicts the 
symmetry of the family members mass matrices, presented in Eq.~(\ref{M0}). It also
offers the explanation for the charges of the family members, for the appearance of the
gauge fields to the family members charges, for the appearance of the dark matter and
for the matter-antimatter asymmetry. For better explanation we encourage the reader to 
read Refs.~\cite{norma2014MatterAntimatter,N2014scalarprop}.

There are, namely, two (only two) kinds of the anticommuting Clifford algebra objects: 
The Dirac $\gamma^a$ take care of the spin in $d=(3+1)$, the spin in  $d\ge 4$ 
(rather than the total angular momentum) manifests in $d=(3+1)$ in the low energy regime
as the charges.  In this part the {\it spin-charge family} theory is like the Kaluza-Klein
theory~\cite{kk,zelenaknjiga}, unifying spin
and charges, and offering a possible answer to the question about the origin of the so far 
observed charges. It also explains why left handed family members are weak charged, while 
the right handed are weak chargeless.

The second kind of the Clifford algebra objects, forming the equivalent representations with 
respect to the Dirac kind, recognized by one of the authors (SNMB), is responsible for the 
appearance of families of fermions. 

There are correspondingly also two kinds of gauge fields, which manifest in $d=(3+1)$ as
the vector gauge fields and as the scalar gauge fields. The vector gauge fields explain 
properties of all known vector gauge fields. The scalar gauge fields with the space index $s$, 
$5\leq s$ $ \leq 8$, offer, as weak doublets carrying also appropriate hyper charge, 
explanations for the appearance of the Higgs scalar and the Yukawa coupling. Those 
scalars with the space index $s$, 
$9 \leq s $  $\leq 14$, are colour triplets causing transitions of antileptons into quarks 
and antiquarks into quarks and back. In the presence of the scalar condensate, 
which breaks  the matter-antimatter symmetry, offer these scalars the explanation for the observed 
matter-antimatter asymmetry,  explaining also the proton decay. All the scalar fields carry 
besides the charges, determined by the space index, also additional charges, which all are 
in the adjoint representations.  

In the {\em spin-charge-family} theory originate all the properties of at low energies 
observed fermions and bosons in a simple starting action for massless fields in
 $d=[1 +(d-1)]$.  The theory makes a choice of $d=(1+13)$, since one Weyl
 representation of $SO(1,13)$ contains all the family members, left and right handed, 
with antimembers included. 

Fermions interact (Eq.~(\ref{action})) with the vielbeins $f^{\alpha}{}_a$ and 
 also  with the two kinds of the spin connection fields: with 
$\omega_{abc}= f^{\alpha}{}_{c}\,$ $\omega_{ab \alpha}$, which are  
the gauge fields of $S^{ab}= \frac{i}{4}\,(\gamma^a \gamma^b - $
$\gamma^b \gamma^a)$, and  with $\tilde{\omega}_{abc} =$ 
$f^{\alpha}{}_{c}\, \tilde{\omega}_{ab \alpha}$, which are the gauge fields of 
$\tilde{S}^{\tilde{a}\tilde{b}}= \frac{i}{4}\, (\tilde{\gamma}^{\tilde{a}}
\tilde{\gamma}^{\tilde{b}} - \tilde{\gamma}^{\tilde{b}}
 \tilde{\gamma}^{\tilde{a}})$. 
 $\alpha, \beta,\dots$ is the Einstein index and 
$a,b,\dots$ is the flat index. The starting action is the simplest one
\begin{eqnarray}
\label{action}
S            \, = \int \; d^dx \; E\;{\mathcal L}_{f} &+&    
                \int \; d^dx \; E\; (\alpha \,R + \tilde{\alpha} \, \tilde{R})\,,
\nonumber\\             
{\mathcal L}_f  = &&\frac{1}{2} (\bar{\psi} \, \gamma^a p_{0a} \psi) + h.c.\,
 \nonumber\\
p_{0a }         = f^{\alpha}{}_a \, p_{0\alpha} + \frac{1}{2E}\, \{ p_{\alpha},\,
 E f^{\alpha}{}_a\}_- \,,
&& \,  
p_{0\alpha}     =  p_{\alpha}  -  \frac{1}{2}  S^{ab}\, \omega_{ab \alpha} - 
                    \frac{1}{2}  \tilde{S}^{\tilde{a} \tilde{b}}  \,
           \tilde{\omega}_{\tilde{a}\tilde{b} \alpha}\,,\nonumber\\                   \\ 
R               =  \frac{1}{2} \, \{ f^{\alpha [ a} f^{\beta b ]} 
\;(\omega_{a b \alpha, \beta} 
- \omega_{c a \alpha}\,\omega^{c}{}_{b \beta}) \} + h.c.\,,&&\,   
\tilde{R}       = \frac{1}{2}\,   f^{\alpha [ \tilde{a}} f^{\beta \tilde{b} ]} \;
(\tilde{\omega}_{\tilde{a} \tilde{b} \alpha,\beta} - 
\tilde{\omega}_{\tilde{c} \tilde{a} \alpha}\, 
\tilde{\omega}^{\tilde{c}}{}_{\tilde{b} \beta}) +
 h.c.\,. \nonumber\\
\end{eqnarray}
$ E = \det(e^a{\!}_{\alpha}) $ and $e^a{\!}_{\alpha} f^{\beta}{\!}_a =$ 
$ \delta^{\beta}_{\alpha} $. The vielbeins $ f^{\alpha}{\!}_{ \tilde{a}}$  $=
f^{\alpha}{\!}_{a}$ (\cite{N2014scalarprop}).

Fermions, coupled to the vielbeins and the two kinds of the spin connection fields,
 {\it  manifest} (after several breaks and the appearance of the scalar condensate 
of the two right handed neutrinos~\cite{norma2014MatterAntimatter,N2014scalarprop,%
JMP,hn}) {\it before the electroweak break four coupled massless families of 
quarks and leptons}, the left handed fermions are weak charged and the right
handed ones are weak chargeless, explaining all the assumptions of the {\it standard 
model}. 

The vielbeins and the two kinds of the spin connection fields manifest effectively as the observed 
gauge fields and (those with the scalar indices with respect to $d=(1+3)$) as  several scalar fields. 
The mass matrices of the four family members (quarks and leptons) are after the 
electroweak break expressible on a tree level by the vacuum expectation values of
 the two kinds of the spin connection fields and the corresponding vielbeins with the 
scalar indices $s=(7,8)$ (\cite{NBled2013,NBled2012,pikanorma,JMP,NPLB,%
N2014scalarprop}):\\
{\bf i.} One kind, originating in the scalar fields $\tilde{\omega}_{abc}\,$, manifests   
as  the two $SU(2)$ triplets -- $\tilde{A}_{s}^{\tilde{N}_L \,i}, i=(1,2,3)\,,s=(7,8)$; 
$\tilde{A}^{\tilde{1}\,i}_{s}\,, i=(1,2,3)\,,s=(7,8)$; --  and one singlet -- 
$\tilde{A}_{s}^{\tilde{4}}\,, s=(7,8)$ -- contributing equally to all the family members. \\
{\bf ii.} The second kind originates in the  scalar fields $\omega_{abc}$, manifesting 
as 
the  three singlets  -- $A^{Q}_{s}, A^{Q'}_s, A^{Y'}\,, s=(7,8)$ ($Q$ is the 
electromagnetic charge, $Q'$ is the non conserved charge of the $Z$ boson, $Y'$ 
originates  in the second $SU(2)$, which breaks at the appearance of the condensate,
leaving the hyper charge $Y$ conserved) -- contributing the same values to all 
 the families and distinguishing among family members.

The scalar fields with $s=(7,8)$ "dress" the right handed quarks and leptons with the
hyper charge and the weak charge so that they manifest the charges of the left 
handed partners, explaining assumptions for the Higgs's scalar of the
 {\it standard model}~\footnote{It is the term $\gamma^0 \gamma^s \,
\phi^{Ai}_s\,$, where $\phi^{Ai}_s$, with $s=(7,8)$, denotes any of the scalar fields,
 which transforms the right handed fermions into the corresponding left handed 
partner~\cite{NBled2013,JMP, NBled2012,%
pikanorma,scalars011,N2014scalarprop,norma2014MatterAntimatter}. This mass
 term originates in $\bar{\psi} \, \gamma^a p_{0a} \psi$ of the 
action~Eq.(\ref{action}), with $a= s=(7,8)$ and $p_{0s}= 
f^{\sigma}_{s}\, (p_{\sigma} - 
\frac{1}{2} \tilde{S}^{ab}\tilde{\omega}_{\tilde{a}\tilde{b} \sigma} - 
\frac{1}{2} S^{st}\omega_{st \sigma})$.}
and contributing also to the masses of the weak bosons,  as doublets with respect to
the weak charge.

Loop corrections, to which all the scalar and also gauge vector fields contribute 
coherently, change contributions of the off-diagonal and diagonal elements appearing 
on the tree level, keeping the tree level symmetry of mass matrices 
unchanged~\footnote{It can be seen that all the loop corrections keep the starting 
symmetry of the mass matrices unchanged in the massless basis. We have also 
started~\cite{JMP,AN} with the evaluation of  the loop 
corrections to the 
tree level values.   
This estimation has been done so far~\cite{AN} only up to the  first order and 
partly to the second order.}.

\subsection{Symmetries of the mass matrices on the tree level and beyond
 manifesting the  $SU(2)\times SU(2) \times U(1)$ symmetry}
\label{M0SCFT}

Let us make a choice of a massless basis $\psi^{\alpha}_{i}$, $i=(1,2,3,4)$, for a particular
family member $\alpha =(u,\nu,d,e)_{L,R}$. The two $SU(2)$ 
operators~\footnote{The infinitesimal generators
$\tilde{N}^{i}_{L}\,,\,i=(1,2,3)$ and $\tilde{N}^{i}_{R}\,,\,i=(1,2,3)$  determine the algebra 
of the two invariant subgroups of the $\widetilde{SO}(1,3)$ group, while 
$\tilde{\tau}^{1i}\,,\,i=(1,2,3)$ and $\tilde{\tau}^{2i}\,,\,i=(1,2,3)$ determine the two invariant 
subgroups of the $\widetilde{SO}(4)$ group. The four families, discussed in this paper, 
carry family quantum numbers of $\tilde{N}^{i}_{L}$ and $ \tilde{\tau}^{1i}$.}, 
$\tilde{N}^{i}_{L}$ and $ \tilde{\tau}^{1i}$,

\begin{eqnarray}
\label{taunl}
\tilde{N}^{i}_{L}\,,\,i=(1,2,3)\,, && {\tau}^{i}_{L}\,,\,i=(1,2,3)\,,
\nonumber\\ 
\{\tilde{N}^{i}_{L}, \tilde{N}^{j}_{L}\}_{-}= i\,\varepsilon^{ijk} \tilde{N}^{k}_{L}\,,\;\;
\{\tilde{\tau}^{1i},\tilde{\tau}^{1j}\}_{-}&=& i\,\varepsilon^{ijk} \tilde{\tau}^{1k}
\,,\{\tilde{N}^{i}_{L}, \tilde{\tau}^{1j}\}_{-}=0\,,
\end{eqnarray}
$\varepsilon^{ijk}$ is the totally antisymmetric tensor, transform the basic 
vectors $\psi^{\alpha}_{i}$, into one another  as follows
\begin{eqnarray}
\label{taunlonpsi}
&&\tilde{N}^{3}_{L}\, (\psi^{\alpha}_1, \psi^{\alpha}_2,\psi^{\alpha}_3,\psi^{\alpha}_4)=
 \frac{1}{2}\, (-\psi^{\alpha}_1, \psi^{\alpha}_2,-\psi^{\alpha}_3,\psi^{\alpha}_4)\,,\nonumber\\
&&\tilde{N}^{+}_{L}\, (\psi^{\alpha}_1, \psi^{\alpha}_2,\psi^{\alpha}_3,\psi^{\alpha}_4)= 
 (\psi^{\alpha}_2, \;\;0,\psi^{\alpha}_4,\;\;0)\,,\nonumber\\
&&\tilde{N}^{-}_{L}\, (\psi^{\alpha}_1, \psi^{\alpha}_2,\psi^{\alpha}_3,\psi^{\alpha}_4)= 
 (0\;\;, \psi^{\alpha}_1,\;\;0,\psi^{\alpha}_3)\,,\nonumber\\
&&\tilde{\tau}^{13}\, (\psi^{\alpha}_1, \psi^{\alpha}_2,\psi^{\alpha}_3,\psi^{\alpha}_4)=
 \frac{1}{2}\, (-\psi^{\alpha}_1, -\psi^{\alpha}_2,\psi^{\alpha}_3,\psi^{\alpha}_4)\,, \nonumber\\
&&\tilde{\tau}^{1+}\, (\psi^{\alpha}_1, \psi^{\alpha}_2,\psi^{\alpha}_3,\psi^{\alpha}_4)= 
(\psi^{\alpha}_3, \psi^{\alpha}_4,\;\;0,\;\;0)\,,\nonumber\\ 
&&\tilde{\tau}^{1-}\, (\psi^{\alpha}_1, \psi^{\alpha}_2,\psi^{\alpha}_3,\psi^{\alpha}_4)=
 (\;\;0, \;\;0,\psi^{\alpha}_1,\psi^{\alpha}_2) \,.
\end{eqnarray}
The three $U(1)$ operators ($Q, Q'$ and $Y'$) commute with the family operators 
$\vec{\tilde{N}}_{L}$ 
and $\vec{\tilde{\tau}}^{1}$, distinguishing only among family members $\alpha$
\begin{eqnarray}
\label{QQ'Y'nltau}
&&\{ \tilde{N}^{i}_{L}\,, (Q,Q',Y') \}_{-} =(0,0,0)\,,\nonumber\\
&&\{ \tilde{\tau}^{1i}\,, (Q,Q',Y') \}_{-} =(0,0,0)\,,\nonumber\\
&&(Q\,, Q'\,, Y')\, (\psi^{\alpha}_1, \psi^{\alpha}_2,\psi^{\alpha}_3,\psi^{\alpha}_4)= 
(Q^{\alpha}\,, Q'^{\alpha}\,, Y'^{\alpha} )\,
 (\psi^{\alpha}_1, \psi^{\alpha}_2,\psi^{\alpha}_3,\psi^{\alpha}_4) \,,
\end{eqnarray}
giving the same eigenvalues for all the families.

The nonzero vacuum expectation values of the gauge scalar fields of
 $\tilde{N}^{i}_{L}$  ($\tilde{A}^{\tilde{N}_{L}i}_{s}=$ $ 
\tilde{C}^{\tilde{N}_{L}i}{}_{\tilde{m}\tilde{n}}\, 
\tilde{\omega}^{\tilde{m}\tilde{n}}{}_{s}$, $(\tilde{m}, \,\tilde{n}) =(0,1,2,3)$), 
 of  $\tilde{\tau}^{1i}$  ($\tilde{A}^{\tilde{1}i}_{s}= 
\tilde{C}^{\tilde{1}i}{}_{\tilde{t} \tilde{t'}}\, \tilde{\omega}^{\tilde{t}\tilde{t'}}{}_{s}
$,  $(\tilde{t},\tilde{t'}) =(5,6,7,8)$) and of the three singlet gauge  scalar fields of  
($Q,Q'$ and $Y'$), which all are 
superposition of $\omega_{t,t',s}$ ($A^{Q}_{s} = C^{Q}{}_{t t'}\,\omega^{t t'}{}_{s}$\,,
 $A^{Q'}_{s} = C^{Q'}{}_{t t'}\,\omega^{t t'}{}_{s}$ and $A^{Y'}_{s} = C^{Y'}{}_{t t'}\,
\omega^{t t'}{}_{s}$, \,$(s,t,t') =(5,6,7,8)$), determine on the tree level, together with the 
corresponding coupling  constants,  the
 $SU(2) \times SU(2) \times U(1)$ symmetry and the strength of the mass matrix of each 
family member $\alpha$, Eq.~(\ref{M0}). 
In loop corrections all the scalar fields - 
 $ \tilde{A}^{\tilde{N}_{L}i}_{s}$,  $\tilde{A}^{\tilde{1}i}_{s}$, 
$A^{Q}_{s}, A^{Q'}_{s},A^{Y'}_{s}$ - contribute to all the matrix elements, 
keeping the symmetry unchanged, Eq.~(\ref{M0}). The twice  two zeros on the tree level 
obtain in loop corrections the value $b$. 

One easily checks that a change of the phases of the left and the right
 handed members, there are $(2n-1)$ 
possibilities, causes  changes in phases of matrix elements in Eq.~(\ref{M0}). 
 
 All the scalars are doublets with respect to the weak charge,  
contributing to the weak and the hyper charge of the fermions so that they
 transform the right handed members into 
the left handed onces~\cite{N2014scalarprop}, what is in the {\it standard model} just
required.

\section{Properties of non Hermitian mass matrices}
\label{nonhermitean}

This pedagogic presentation of well known properties of non Hermitian matrices can be 
found in many textbooks, for example~\cite{chine}. We repeat this topic here only to make our discussions  
transparent.

Let us take   a  non hermitian mass matrix  $M^{\alpha}$ as it follows from the 
{\it spin-charge-family} theory, $\alpha$ denotes a family member 
(index ${}_{\pm}$ used in the main text is dropped).

We always can diagonalize  a non Hermitian $M^{\alpha}$ with two unitary matrices, 
$S^{\alpha}$ ($S^{\alpha\, \dagger}\,S^{\alpha}=I$)
and $T^{\alpha}$ ($T^{\alpha\, \dagger}\,T^{\alpha}=I$)
\begin{eqnarray}
\label{diagnonher0}
S^{\alpha\, \dagger}\,M^{\alpha}\,T^{\alpha}&=& {\bf M}^{\alpha}_{d}\,=
(m^{\alpha}_{1}\, \dots m^{\alpha}_{i}\, \dots m^{\alpha}_{n}).
\end{eqnarray}
The proof is added below. 

Changing phases of the basic states, those of the left handed one and those of the right handed one, the 
new unitary matrices  $S'^{\alpha} = S^{\alpha} \,F_{\alpha S} $ and $T'^{\alpha} = T^{\alpha}\,F_{\alpha T}$
change the phase of the elements of diagonalized mass matrices ${\bf M}^{\alpha}_{d}$ 
\begin{eqnarray}
\label{diagnonher}
S'^{\alpha\, \dagger}\,M^{\alpha}\,T'^{\alpha}&=& F^{\dagger}_{\alpha S}\,{\bf M}^{\alpha}_{d}\,
F_{\alpha T}=\nonumber\\
& & diag(m^{\alpha}_{1} e^{i(\phi^{\alpha S}_{1}- \phi^{\alpha T}_{1})}\, \dots 
m^{\alpha}_{i}\, e^{i(\phi^{\alpha S}_{i}- \phi^{\alpha T}_{i})}\,, \dots m^{\alpha}_{n}\,
e^{i(\phi^{\alpha S}_{n}- \phi^{\alpha T}_{n})})\,,\nonumber\\
F_{\alpha S} &=& diag(e^{-i \phi^{\alpha S}_{1}},\,\dots\,,e^{-i \phi^{\alpha S}_{i}}\,,\dots\,,
e^{-i \phi^{\alpha S}_{n}})\,,\nonumber\\
F_{\alpha T} &=& diag(e^{-i \phi^{\alpha T}_{1}},\,\dots\,,
e^{-i \phi^{\alpha T}_{i}}\,,\dots\,, e^{-i \phi^{\alpha T}_{n}})\,.
\end{eqnarray}

In the case that the mass matrix is Hermitian $T^{\alpha}$ can be replaced by $ S^{\alpha}$, but only up 
to phases originating in the phases of the two basis, the left handed one and the right handed one, 
since they remain independent. 

One can diagonalize the non Hermitian mass matrices in two ways, that is either one diagonalizes  
$M^{\alpha }M^{\alpha \,\dagger}$ or $M^{\alpha \dagger} M^{\alpha }$
\begin{eqnarray}
\label{diagMM}
(S^{\alpha \dagger} M^{\alpha} T^{\alpha})          (S^{\alpha \dagger} M^{\alpha } T^{\alpha})^{\dagger}&=&
S^{\alpha \dagger} M^{\alpha } M^{\alpha \,\dagger} S^{\alpha} = {\bf M}^{\alpha  2}_{d S}\,, \nonumber\\  
(S^{\alpha \dagger} M^{\alpha} T^{\alpha})^{\dagger}(S^{\alpha \dagger} M^{\alpha } T^{\alpha})&=&
T^{\alpha \dagger} M^{\alpha\, \dagger} M^{\alpha } T^{\alpha} = {\bf M}^{\alpha  2}_{d T}\,, \nonumber\\
{\bf M}^{\alpha\, \dagger }_{d S}&=& {\bf M}^{\alpha }_{d S}\, , 
\quad {\bf M}^{\alpha\, \dagger }_{d T}= {\bf M}^{\alpha}_{d T}\,.
\end{eqnarray}
One can prove that ${\bf M}^{\alpha }_{d S}={\bf M}^{\alpha }_{d T}$. The proof proceeds as follows.
Let us define two Hermitian ($H^{\alpha}_{S}\,$, 
$H^{\alpha}_{T}$) and two unitary matrices ($U^{\alpha}_{S}\,$, $H^{\alpha}_{T}$) 
\begin{eqnarray}
\label{proof1}
H^{\alpha}_{S} &=& S^{\alpha} {\bf M}^{\alpha }_{d S} S^{\alpha \, \dagger}\,, \quad \quad
H^{\alpha}_{T}  =  T^{\alpha} {\bf M}^{\alpha \dagger }_{d T} T^{\alpha \, \dagger}\,,\nonumber\\
U^{\alpha}_{S} &=& H^{\alpha -1}_{S} M^{\alpha } \,, \quad \quad U^{\alpha}_{T} =
 H^{\alpha -1}_{T} 
M^{\alpha \,\dagger } \,.
\end{eqnarray}
It is easy to show  that $H^{\alpha\, \dagger}_{S}= H^{\alpha}_{S}$, 
$H^{\alpha\, \dagger}_{T}= H^{\alpha}_{T}$, $U^{\alpha}_{S} \,U^{\alpha\, \dagger}_{S}=I$ and 
$U^{\alpha}_{T} \,U^{\alpha\, \dagger}_{T}=I$.
Then it follows  
\begin{eqnarray}
\label{proof2}
S^{\alpha \dagger}\, H^{\alpha}_{S} \,S^{\alpha} &=& {\bf M}^{\alpha }_{d S}= {\bf M}^{\alpha \,\dagger}_{d S}=
S^{\alpha \dagger}\,M^{\alpha }\,U^{\alpha \,-1}_{S} \,S^{\alpha}= S^{\alpha \dagger}\,M^{\alpha }\, T^{\alpha} \,,
\nonumber\\
T^{\alpha \dagger}\, H^{\alpha}_{T} \,T^{\alpha} &=& {\bf M}^{\alpha }_{d T}= {\bf M}^{\alpha \,\dagger}_{d T}=
T^{\alpha \dagger}\,M^{\alpha \, \dagger}\,U^{\alpha \,-1}_{T} \,T^{\alpha}= T^{\alpha \dagger}\,
M^{\alpha \dagger}\, S^{\alpha} \,,
\end{eqnarray}
where we recognized $U^{\alpha \,-1}_{S} \,S^{\alpha}= T^{\alpha}$ and $U^{\alpha \,-1}_{T} \,T^{\alpha}=S^{\alpha}$. 
Taking into account Eq.~(\ref{diagnonher}) the starting basis can be chosen so, that all diagonal 
masses are real and positive.

\section*{Acknowledgments} The author acknowledges funding of the Slovenian Research Agency.


\begin{thebibliography}{99}
\bibitem{NPLB} N.S. Manko\v c Bor\v stnik, "Do we have the explanation for
the Higgs and Yukawa couplings of the {\em standard model}", [arxiv:1212.3184,
arxiv:1011.5765].
\end{thebibliography}
\end{document}